\documentclass[usenatbib,usegraphicx]{mn2e}

\begin{document}

\title[X-ray Groups of Galaxies in AEGIS]{Groups of Galaxies in AEGIS: The 200 ksec \textit{Chandra} Extended X-ray Source catalogue }

\author[T.E. Jeltema et al.]{Tesla E. Jeltema $^1$, Brian F. Gerke $^2$, Elise S. Laird $^3$, Christopher N. A. Willmer $^4$, 
\newauthor
Alison L. Coil $^5$, Michael C. Cooper $^{4,6}$, Marc Davis $^7$, Kirpal Nandra $^3$, 
\newauthor
Jeffrey A. Newman $^8$\\
 $^1$Morrison Fellow, UCO/Lick Observatories, 1156 High St., Santa Cruz, CA 95064\\
 $^2$Kavli Institute for Particle Astrophysics and Cosmology, SLAC, 25 Sand Hill Rd. MS 29, Menlo Park, CA 94725\\
 $^3$Astrophysics Group, Blackett Laboratory, Imperial College London, Prince Consort Rd, London SW7 2AZ\\
 $^4$Steward Observatory, University of Arizona, 933 North Cherry Avenue, Tucson, AZ 85721\\
 $^5$Department of Physics and Center for Astrophysics and Space Sciences, University of California, San Diego, 9500 Gilman Dr., La Jolla, CA 92093\\
  $^8$Spitzer Fellow\\
 $^7$Department of Astronomy, University of California at Berkeley, Mail Code 3411, Berkeley, CA 94720\\
 $^8$University of Pittsburgh, Physics \& Astronomy Department, 3941 O'Hara Street Pittsburgh, PA 15260}

\maketitle
\begin{abstract}
We present the discovery of seven X-ray emitting groups of galaxies selected as extended X-ray sources in the 200 ksec \textit{Chandra} coverage of the All-wavelength Extended Groth Strip International Survey (AEGIS).  In addition, we report on AGN activity associated to these systems.  For the six extended sources which lie within the DEEP2 Galaxy Redshift Survey coverage, we identify optical counterparts and determine velocity dispersions.  In particular, we find three massive high-redshift groups at $z>0.7$, one of which is at $z=1.13$, the first X-ray detections of spectroscopically selected DEEP2 groups.  We also present a first look at the the $L_X-T$, $L_X-\sigma$, and $\sigma-T$ scaling relations for high-redshift massive groups.  We find that the properties of these X-ray selected systems agree well with the scaling relations of similar systems at low redshift, although there are X-ray undetected groups in the DEEP2 catalogue with similar velocity dispersions.  The other three X-ray groups with identified redshifts are associated with lower mass groups at $z\sim0.07$ and together form part of a large structure or "supergroup" in the southern portion of the AEGIS field.  Similar to other X-ray-luminous groups, all of the low-redshift systems are centred on massive elliptical galaxies, and all of the high-redshift groups have likely central galaxies or galaxy pairs.  Interestingly, the central galaxies in the highest redshift system show indications of ongoing star formation.  All of the central group galaxies host X-ray point sources, radio sources, and/or show optical AGN emission.  Particularly interesting examples of central AGN activity include a bent-double radio source plus X-ray point source at the center of a group at $z=0.74$, extended radio and double X-ray point sources associated to the central galaxy in the lowest-redshift group at $z=0.066$, and a bright green valley galaxy (part of a pair) in the $z=1.13$ group which shows optical AGN emission lines.
\end{abstract}

\section{ INTRODUCTION }

Groups of galaxies represent a more common, more typical environment than their well-studied cousins massive clusters.  Most galaxies in the local universe lie in groups (e.g.~Turner \& Gott 1976) and groups lie at a transition between the densest environments in the universe, cluster cores, and the field, making them important environments in the study of galaxy evolution (e.g.~Zabludoff \& Mulchaey 1998).   Groups are also important environments for diagnosing the origin of the non-gravitational heating of the intracluster medium (ICM).  From low-redshift studies, non-gravitational heating of the ICM appears to be proportionally more important in groups than in clusters, leading to a steepening of the $L_X-T$ relation for group mass systems (e.g.~Helsdon \& Ponman 2000)  and excess group entropy compared to the self-similar expectations based on massive clusters (e.g.~Ponman et al.~2003; Sun et al.~2009).  In fact, the evolution of the group X-ray scaling relations with redshift can help to constrain models of non-gravitational heating (Balogh et al.~2006), distinguishing, for example,  between excess entropy that is constant with redshift (i.e.~preheating) versus entropy that is tied to halo cooling time (i.e.~increasing with time).

Groups of galaxies, and in particular their evolution with redshift, have been relatively poorly studied compared to massive clusters.  Their faint X-ray luminosities and low galaxy densities make groups difficult to detect outside of the local universe, and we have only recently begun to study the evolution of this important environment with redshift (e.g.~Mulchaey et al. 2006; Jeltema et al. 2006, 2007, 2008a; Willis et al. 2005; Pacaud et al. 2007; Wilman et al. 2005a,b; Gerke et al. 2007).  For example, only a handful of groups/poor clusters at moderate redshifts ($0.2<z<0.6$) have both deep enough X-ray data and optical spectroscopy to be able to look at the relationships between group X-ray temperature, luminosity and group velocity dispersion (Jeltema et al. 2006, 2007, 2008a; Mulchaey et al.~2006; Willis et al. 2005; Gastaldello et al. 2008).   

In this paper, we seek to extend studies of the X-ray emission from groups to higher redshifts using the deep \textit{Chandra} data in the Extended Groth Strip (EGS).  These data cover a $\sim0.67$ deg$^2$ field with a nominal exposure time of 200 ksec.  The area covered represents an increase by a factor of eight over a previous search for groups in a single EGS Chandra pointing \citep{fang} which yielded no detections.  The depth of the \textit{Chandra} data allows us to both detect and to determine average temperatures for groups at high redshifts ($z>0.5$).  Specifically, we conduct an X-ray search for groups through their extended X-ray emission.  In addition to the \textit{Chandra} coverage, the EGS field was one of four fields targeted by the DEEP2 spectroscopic survey (Davis et al.~2003), giving the added benefit of deep optical spectroscopic coverage.  In particular, the DEEP2 spectroscopy in the EGS is magnitude limited, but not colour-selected, giving a large galaxy sample at all redshifts and allowing us to optically identify and determine velocity dispersions for X-ray-detected group candidates (Davis et al.~2007).  These data allow us to take a first look at the $L_X-T$, $L_X-\sigma$, and $\sigma-T$ scaling relations for high-redshift groups and poor clusters, and in future work we will investigate the properties of galaxies in these systems.  While previous X-ray surveys have found a few high-redshift groups, these currently lack this depth of optical spectroscopy \citep{fi07,cdfn}.

Our data reduction, group selection and optical identification procedures are presented in \S2 and \S3.  In \S4 we present our main results including the discovery of a "supergroup" at $z\sim0.07$ (\S4.1), three high-redshift groups (\S4.2) at $z>0.7$, and several group AGN (\S4.3).  The high-redshift systems represent the first X-ray detections of DEEP2 spectroscopically selected groups \citep{ge05,fang}, and we present a first look at the scaling relations of systems of this mass at high redshifts.  Our results are summarised in \S5, and in the appendix we discuss the one X-ray-selected group for which no optical identification was possible.  Throughout the paper, we assume a cosmology of $H_0=70h_{70}$ km s$^{-1}$ Mpc$^{-1}$, $\Omega_m=0.27$, $\Lambda = 0.73$ and $E_z = (\Omega_m(1+z)^3 + \Omega_{\Lambda})^{1/2}$.

\section{ \textit{Chandra} DATA AND REDUCTION }

The Extended Groth Strip was observed with a series of eight \textit{Chandra} ACIS-I pointings, one in Cycle 3 (Nandra et al.~2005) and seven in Cycle 6 (Larid et al.~2009), as a part of the All-wavelength Extended Groth Strip International Survey (AEGIS).  The number of observations per ACIS-I pointing varies between three and ten, but all fields have approximately the same total depth of 200 ksec.  Details about the survey and the AEGIS X-ray point source catalogue can be found in Laird et al.~(2009).  Here we present the first extended X-ray source list.

To search for extended sources (see \S3.1) we utilise the combined, soft band (0.5-2 keV) images presented in Laird et al.~(2009), and we refer the reader to this work for a detailed discussion of the reduction procedure.  In brief, these images were created starting from the level 1 event files; known aspect offsets were removed, bad pixels and afterglows were detected and removed, and the CTI and time dependent gain corrections were applied.  The observations taken in Cycle 6 (seven of the eight pointings) were taken in VFAINT mode, and for these observations the additional VFAINT mode background cleaning was applied.  Background flare time periods were removed following the procedure described in Nandra et al.~(2007) and excluding time periods where the background exceeded 1.4 times the quiescent rate.  From these cleaned event files, images and exposure maps were created in several energy bands; here we focus on the soft band (0.5-2 keV) for group detection as this band maximises the signal-to-noise for soft, thermal sources like groups.  Images and exposure maps of overlapping observations were aligned and merged using the centroids of bright sources in the field, creating one merged image per ACIS-I pointing.  For group detection, we utilise these merged images and exposure maps.

For each detected candidate group, we also extract spectra to determine the average source luminosity and temperature (\S3.3).  In the spectroscopic analysis, we wish to consider each observation separately, allowing us to create observation specific spectral response files (RMFs and ARFs).  Here we followed a very similar data reduction path to create individual observation level 2, flare filtered event files to the one outlined above.  However, we employ a slightly more restrictive flare filtering technique matching the prescriptions of Markevitch et al. (2003).  The filtering excluded time periods when the count rate, excluding point sources and group emission, was not within 20\% of the quiescent rate in the 0.3-12 keV band.  In practice, most of the observations have little contamination from background flares and the difference in flare filtering is negligible.  In total, the AEGIS \textit{Chandra} data were divided in to 42 observations; for nearly all of these, the net flare filtered exposure times differ by less than a couple percent and only two observations have net exposure times differing by more than 20\%.  

For each group candidate identified below, we extracted spectra, after the removal of point sources, within the radius where the total group flux was above the background.  Point sources are detected using wavdetect with spatial scales of 1, 2, 4, 8, and 16 pixels.  They were typically excluded using the output wavdetect 3$\sigma$ ellipses, though for some sources embedded within the group emission or where two point sources had a very small separation the source regions were adjusted by hand.  Background spectra were extracted from a large annular region surrounding the source region, adjusting as necessary to avoid chip edges.  We used the CIAO script specextract to create source and background spectra, RMF and ARF files for each observation of a given group candidate.  As the total net group counts are typically quite low, we employ the fitting procedure for low count data described in Willis et al.~(2005) and applied to the \textit{XMM} Large-Scale Structure survey (XMM-LSS).  Each spectrum is binned such that the associated background spectrum has at least five counts per bin.  Spectra are fit within XSPEC (v12.4.0ao) using the C-statistic, which is considerably more accurate than the $\chi^2$ statistic for data with low counts per bin (Cash 1979).  The C-statistic within XSPEC (cstat) is slightly modified to account for statistical fluctuations in the background, and the slight binning of the data and backgrounds employed here alleviates biases in the model parameters that can occur when many of the spectral bins contain zero background counts \citep{will}.

\section{ GROUP SELECTION AND DETERMINATION OF GROUP PROPERTIES}

\subsection{ X-ray Selection}

To detect extended X-ray sources, we employ the Voronoi Tessellation and Percolation algorithm (VTP; Ebeling \& Wiedenmann 1993) run on the soft band images and exposure maps.  Specifically, we utilise the CIAO tool VTPDETECT with a false positive threshold of $1 \times 10^{-5}$ and a minimum of 20 source counts.  To consider a source as extended, we required that the average of the VTPDETECT ellipse minor and major axes exceed three times the 95\% encircled energy radius at the source off-axis distance and that the source have a signal-to-noise $> 3$ once nearby point sources were removed.  Candidate extended sources were then examined by eye to confirm their extended nature; in particular, closely spaced point sources can lead to false detections.  We also examined by eye the full (0.5-7 keV) and hard (2-7 keV) band images to search for possible contamination from hard point sources that may be faint in the soft band images.  A similar methodology was used to find groups and clusters in the CDF-N \citep{cdfn} and previously in the single \textit{Chandra} pointing in the EGS taken in Cycle 3 \citep{fang}.  In total, our search for extended X-ray sources yielded seven candidate groups, listed in Table 1.  

We detect extended sources to an observed flux limit of about $2 \times 10^{-15}$ ergs cm$^{-2}$ s$^{-1}$.  All of the detected sources are found in regions of fairly high effective exposure ($\geq 100$ ksec), giving a survey area of $\sim 0.53$ deg (see Figure 2 in Laird et al.~(2009)) and an extended source density of 13.2 deg$^{-2}$.  Given the small number statistics and differences in selection technique this density is consistent within the errors with previous surveys.  For example, the Chandra Deep Fields North and South each detect one extended source near our flux limit in a single ACIS-I pointing \citep{cdfn,cdfs}, while the the XMM COSMOS survey \citep{fi07} and projection of the ROSAT Deep Cluster Survey \citep{rdcs} to lower fluxes give somewhat higher densities of 20-35 deg$^{-2}$ at our flux limit.

\subsection{ Optical Identification}

To identify potential optical group counterparts for the extended X-ray sources,
we constructed a catalogue of galaxy groups in the EGS using the DEEP2
spectroscopic data in the AEGIS region.  Unlike the other three DEEP2 fields where high-redshift ($z>0.7$) galaxies were pre-selected based on colour for spectroscopic follow-up, in AEGIS, galaxies of all redshifts were targeted for spectroscopy \citep{deep}, allowing us to search for groups at $z<1.5$.  We used the Voronoi-Delaunay method (VDM) group-finder
described in Gerke et al. (2005), with the same parametrization described
in that paper.  This group-finder identifies groups by searching for galaxy
overdensities in redshift space; our initial group catalogue used only
galaxies with good (quality 3 or 4) DEEP2 redshifts.  We then searched this
catalogue for groups that fall near each X-ray source on the sky.  By
inspecting  these groups' member galaxies in the imaging data, and comparing
to the X-ray contours, we identified a best-match optical counterpart for
each X-ray source.  The group-finder parametrization is optimised for the galaxy sampling rate of the overall DEEP2 survey, which is less dense than in EGS.  The optical group catalogue may, therefore, contain false-positives or interloper contamination.  However, here we are simply using this catalogue as a tool to match already identified X-ray groups, and as described in \S3.3, we refine the selection of group member galaxies when determining the velocity dispersions.

The best optical group matches to each extended X-ray source are shown in Figures 1 and 2 compared to the \textit{Chandra} soft band X-ray contours.  Here we show both the contours of the diffuse X-ray emission (in black) and X-ray point sources (in green).  For the purposes of illustration, the diffuse contours were created by first filling point source regions (from wavdetect) using the CIAO tool dmfilth through Poisson sampling with the distribution mean determined from a local background region.  The filled image is then adaptively smoothed with the tool csmooth and exposure corrected.  One of our X-ray candidates, Group 3, unfortunately falls outside of the DEEP2 spectroscopic coverage and no optical identification was possible.  We discuss this group in more detail in the Appendix.  

Three of the extended X-ray sources are identified as low-redshift, low X-ray luminosity groups ($z<0.1$; Figure 1).  In particular, the three low-redshift systems form part of a superstructure or ``supergroup'' at $z \sim 0.07$ in the southern half of the AEGIS field.  The remaining three group candidates are likely high-redshift, X-ray bright groups ($z>0.7$; Figure 2).  These three systems represent the first X-ray confirmation of spectroscopically selected, high-redshift DEEP2 groups.  In one case, Group 5, two DEEP2 groups are good matches to the X-ray source position, one at $z=0.737$ (Figure 2) and one at $z=0.201$ (Figure 3).  Here we consider the high-redshift system to be a better match both in terms of the location of the group members, including the likely central galaxy, and the average velocity dispersion and inferred X-ray luminosity.  As shown in Figure 3, the galaxies in the $z=0.201$ group are offset with respect to the X-ray contours, and this group has a fairly low velocity dispersion of 232 km s$^{-1}$.  However, the X-ray emission does show an extension in the direction of three members of the $z=0.201$ system, one bright spiral galaxy with two smaller companions, indicating that the X-ray emission may be partially due to this foreground group.

\subsection{ Determination of Group Properties }

For each identified group, we determine the average X-ray luminosity and temperature from the \textit{Chandra} spectra.  We fit the spectra from each observation of a given group candidate jointly to an absorbed thermal plasma model (mekal) with the absorption fixed at the Galactic value \citep{nh}, which ranges from $1.0-1.2 \times 10^{20}$ cm$^{-2}$, and the metallicity fixed at 0.3 solar \citep{abun}.  In most cases, the spectra are fit in the $0.5-5.0$ keV band.  In all but one case, we are able to constrain the X-ray temperature.  Group 2 is unfortunately very faint.  For this source, we restrict the spectrum to the 0.5-2 keV range to increase the signal-to-noise and quote the best-fitting luminosity for a fixed temperature of 1.5 keV; we include the uncertainty in the temperature in the luminosity errors by varying the temperature between 0.5 and 10 keV.  
The average temperatures and X-ray luminosities of each group within the spectral extraction radius ($r_{spec}$) are listed in Table 2.  For Group 3, we do not know the group redshift, but the spectrum still allows us to estimate the temperature.  For this group, we left the redshift as a free parameter in the spectral fit when determining the temperature and temperature errors.  The best-fitting temperature is not highly redshift dependent.  We confirmed that leaving the redshift free gave reasonable results by refitting with a fixed redshift which we varied between 0.05 and 1.0; the best-fitting temperature was always within the errors quoted in Table 2.

In order to compare to other group samples, we wish to compare the luminosities within a similar radius.  Therefore, in the scaling relations presented in \S4, we project the group luminosities to $r_{500}$, the radius at which the group density is 500 times the critical density.  We estimate $r_{500}$ from the average temperature using the relation $r_{500} = 0.391 T^{0.63} E_z^{-1}$ Mpc (Table 2; Willis et al.~2005; Finoguenov, Reiprich, \& B\"{o}hringer 2001).  We extrapolate the luminosities from the spectral extraction radius based on a $\beta$-model surface brightness distribution with $\beta = 2/3$ and $r_c = 180$ $h_{70}^{-1}$ kpc, typical values found for clusters and X-ray luminous intermediate-redshift groups (e.g.~Jones \& Forman 1999; Jeltema et al.~2006).
The high-redshift AEGIS groups are detected to at least half of $r_{500}$, and therefore, the extrapolation in luminosity is not large (see Table 2).  However, the luminosities of the low-redshift groups change by factors of $8-25$ due to the small detection radii.  Given the size of this correction, we choose to be conservative in our estimates of the luminosity errors.  For all groups (low and high redshift), we include the effect of the uncertainty in $r_{500}$ due to the uncertainty in the group temperature in the projected luminosity errors.  We also bound the error from the unknown $\beta$-model parameters by using $\beta = 0.5$ and $r_c = 50$ $h_{70}^{-1}$ kpc, which may be more appropriate values for low temperature groups \citep{mu03, will, gems}; this effectively sets the lower limits on our projected luminosities.

For each group, we also determine the velocity dispersion of the member galaxies.  In addition to the DEEP2 redshifts, in this step,
we included galaxies with good-quality redshifts from the Sloan Digital
Sky Survey (Adelman-McCarthy et al.~2008), using data from the NYU Value-Added Galaxy catalogue
(Blanton et al. 2005), and galaxies with redshifts that were obtained in
follow-up observations of the DEEP2 sample with the Hectospec spectrograph on
the MMT (Coil et al.~2009; Willmer et al., in preparation).  With the addition of these galaxies and because the group-finder we used was calibrated for the regions of DEEP2
outside of the EGS, we refined
the membership of each optical counterpart "by hand," by searching for
galaxies within a redshift-space cylinder, oriented with its axis along the
line of sight, with radius and length chosen to produce a well-formed group
on the sky, with a reasonably Gaussian velocity distribution.

There is inevitably some arbitrariness in the choice of this redshift-space cylinder, but we have checked explicitly
that our results are not very sensitive to the details of the cylinder we
choose in each case.  In particular, systematic errors on the group velocity
dispersions arising from the choice of cylinder are smaller than the
statistical uncertainties due to discrete sampling of the velocity
distribution.  We computed these statistical uncertainties using the same
procedure as in Fang et al. (2007).  In brief, we use Monte-Carlo
simulations to determine the probability distribution $P(\sigma_m |
\sigma_t, N)$ of measured velocity dispersions $\sigma_m$, given a true
dispersion $\sigma_t$ and a number of samples $N$.  One can then use Bayes's
theorem to derive the likelihood distribution $\mathcal{L}(\sigma_t|\sigma_m,N)$ of $\sigma_t$ given an observed
$\sigma_m$ and $N$, and our error bars on the group velocity dispersion
reflect the 68\% confidence region of this distribution. The on-sky
distributions of galaxies in the optical-counterpart groups are shown in
Figures 1 and 2, and the velocity dispersions and memberships of the groups are compiled in Table
1.

\onecolumn
\begin{figure}
\caption{ Low-redshift groups in AEGIS showing extended X-ray emission.  From top to bottom these are: Group 4 at $z=0.066$,  Group 2 at $z=0.075$, and Group 1 at $z=0.074$.  The left panels show the \textit{Chandra} 0.5-2 keV contours overlaid on the CFHT R-band images; images are 300 $h_{70}^{-1}$ kpc on a side.  Contours of diffuse X-ray emission, from an adaptively smoothed image with point sources removed, are shown in black, and X-ray point sources are shown in green.  Galaxies identified as group members are marked with red circles.  The right panel shows the spatial distribution of group member galaxies.  The circle areas are proportional to the galaxy's K-corrected optical luminosity (g$_{0.1}$ for SDSS galaxies and Johnson B otherwise) relative to $L_{\ast}$ at the group redshift, and they are colour coded by velocity relative to the central group velocity (determined individually for each group).  Dashed circles show the group radii assumed in the calculation of the velocity dispersion.  Other galaxies in the field (with DEEP2, MMT, or SDSS spectroscopy) within 1500 km s$^{-1}$ of any of the groups are shown as grey triangles. The very small data points show all galaxies in the field with DEEP2 redshifts to indicate the extent of the DEEP2 spectroscopic coverage; redshifts outside this area are taken from SDSS or MMT. }
\end{figure}

\begin{figure}
\caption{ High-redshift groups in AEGIS showing extended X-ray emission.  From top to bottom these are: Group 5 at $z=0.737$, Group 6 at $z=0.766$, and Group 7 at $z=1.132$.  Contours and plot points are the same as in Figure 1, but the X-ray/CFHT overlays are 1 $h_{70}^{-1}$ Mpc on a side and the galaxy distributions are shown for each group individually. Other galaxies in the field (with DEEP2, MMT, or SDSS spectroscopy) within 1500 km s$^{-1}$ of each group are shown as grey triangles.}
\end{figure}

\twocolumn
\begin{figure}
\begin{center}
\includegraphics[width=70mm]{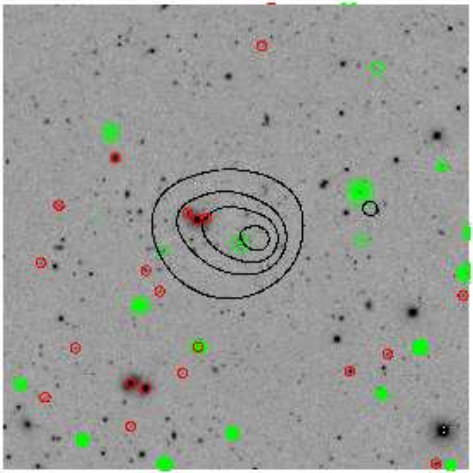}
\end{center}
\caption{  Extended X-ray source Group 5 compared to the $z=0.201$ group in the field.  The X-ray emission shows an extension to the northeast in the direction of three members of the $z=0.201$ system, but the group as a whole is offset to the east of the X-ray emission.  The image is 1 $h_{70}^{-1}$ Mpc comoving on a side for $z=0.201$. }
\end{figure}

\begin{table*}
\begin{minipage}{135mm}
\caption{ Group Candidates and Optical Identifications }
\begin{tabular}{lcccccc}
\hline
Source & RA & Dec & z & $\sigma_v$ (km s$^{-1}$) & $N_{gals}$ & Notes \\
\hline
1 & 14:15:03.16 &+52:04:40.2 &0.074 &$454^{+96}_{-38}$ &34 & \\ %6
2 & 14:15:20.14 &+52:20:48.3 &0.075 &$288^{+72}_{-40}$ &23 &previously identified in SDSS \\ %5
3 & 14:16:16.63 &+52:06:01.9 &- &- &- &insufficient spectroscopy\\ %1
4 & 14:17:49.60 &+52:41:44.0 &0.066 &$404^{+107}_{-53}$ &18 & \\ %8
5 & 14:20:33.66 &+53:08:25.5 &0.737 &$464^{+128}_{-61}$ &17 &possibly $z=0.201$\\ %4
6 & 14:22:39.23 &+53:21:35.1 &0.766 &$367^{+94}_{-45}$ &20 & \\ %7
7 & 14:23:41.67 &+53:28:28.9 &1.129 &$702^{+291}_{-150}$ &8 &  \\%2
\hline
\end{tabular}

\medskip
Columns 2 and 3 list the position of the X-ray centre for each group candidate.  Column 4 lists the central redshift of the best-matched optically selected group from the DEEP2 catalogue.  The average group velocity dispersion and number of group member galaxies from the combined DEEP2, MMT, and SDSS spectroscopy are listed in columns 5 and 6, respectively.
\end{minipage}
\end{table*}

\begin{table*}
\begin{minipage}{180mm}
\caption{ X-ray Properties }
\begin{tabular}{lcccccccc}
\hline
Group & Name & $r_{spec}$ & Spectral & kT & $L_{X,0.5-2}$ & $L_{X,bol}$ & $r_{500}$ & Extrap. \\
& & $h_{70}^{-1}$ kpc & Counts & keV & $10^{42} h_{70}^{-2}$ ergs s$^{-1}$ & $10^{42} h_{70}^{-2}$ ergs s$^{-1}$ & $h_{70}^{-1}$ kpc & Factor \\
\hline
1&CXOU J141503.1+520440 &62 &216 &$0.63^{+0.11}_{-0.11}$ &$0.0736^{+0.0135}_{-0.0119}$ &$0.126^{+0.023}_{-0.020}$ &283 &8.4 \\ %6
2 &CXOU J141520.1+522048 &42$^{\dagger}$ &37 &\textit{1.5} &$0.0492^{+0.0365}_{-0.0174}$ &$0.107^{+0.170}_{-0.037}$ &\textit{489} &24.9 \\ %5
3 &CXOU J141616.6+520601 &(25'')  &224 &$2.89^{+1.04}_{-0.50}$ &- &- &- &-\\ %1
4 &CXOU J141749.6+524144 &62 &156 &$1.50^{+0.50}_{-0.25}$ &$0.0471^{+0.0091}_{-0.0090}$ &$0.103^{+0.020}_{-0.020}$ &491 &11.9 \\ %8
5 &CXOU J142033.6+530825 &364 &260 &$2.34^{+1.10}_{-0.57}$ &$14.9^{+2.6}_{-2.6}$ &$37.3^{+6.8}_{-6.4}$ &456 &1.1 \\ %4
6 &CXOU J142239.2+532135 &259 &71 &$2.00^{+3.20}_{-0.74}$ &$7.87^{+2.79}_{-2.68}$ &$19.0^{+6.8}_{-6.5}$ &406 &1.4 \\ %7
7 &CXOU J142341.6+532828 &247 &67 &$3.58^{+4.59}_{-1.53}$ &$14.4^{+6.6}_{-4.2}$ &$47.4^{+21.7}_{-13.9}$ &477 &1.6 \\ %2
\hline
\end{tabular}

\medskip
Columns 3 and 4 list the X-ray spectral extraction radius and the net spectral counts for each group.  Columns 5-7 list the average group temperature and luminosity (soft band and bolometric) within $r_{spec}$.  Column 8 lists $r_{500}$ calculated as described in \S3.3, and the final column lists the multiplicative factor applied to extrapolate the measured luminosities to $r_{500}$.  Values in italics indicate that the temperature was fixed at 1.5 keV.  $\dagger$ For Group 2 the spectral extraction radius was reduced with respect the the total group extent to avoid a chip edge present in some of the observations.
\end{minipage}
\end{table*}

\section{ RESULTS }

\subsection{ Low-Redshift Groups }

The three extended X-ray sources associated with low-redshift groups are shown in Figure 1.  These three systems lie in the southern portion of the AEGIS \textit{Chandra} coverage, and they all appear to lie in a superstructure at $z \sim 0.07$.  The galaxy distribution in this superstructure is shown in the right panel of Figure 1.  Two of the groups lie close together in both position and velocity space, while the northern most group lies in the foreground at $z=0.066$.  A couple of other associations of $3-4$ group-scale systems, or supergroups, have been found previously (Gonzalez et al.~2005; Brough et al.~2006), and these systems may be the precursors of massive clusters formed through multiple mergers.  However, despite the obvious large-scale structure in the EGS seen in Figure 1, we find that even the closer two groups (1 and 2) are unlikely to be bound given their projected distance ($\sim 1.4 h_{70}^{-1}$ Mpc) and relative velocity ($\sim 300$ km s$^{-1}$) (Beers, Geller, \& Huchra 1982).

Notable from Figure 1 (left) is the fact that all three groups have luminous elliptical galaxies at their X-ray centres, similar to other X-ray-bright, low-redshift groups \citep{mu03, gems}.  These groups are fairly X-ray faint having measured bolometric X-ray luminosities of $\sim 10^{41}$ $h_{70}^{-2}$ ergs s$^{-1}$ and luminosities within $r_{500}$ of $\sim 10^{42}$ $h_{70}^{-2}$ ergs s$^{-1}$.  The detected X-ray emission in all cases appears to extend beyond the optical extent of the central galaxy, consistent with group-scale emission.  For all three groups the detected emission extends beyond 60 kpc, the criterion used to by Osmond \& Ponman (2004) to separate galactic halos from group-scale halos, as shown by the surface brightness profiles in Figure 4.

\begin{figure}
\includegraphics[width=70mm]{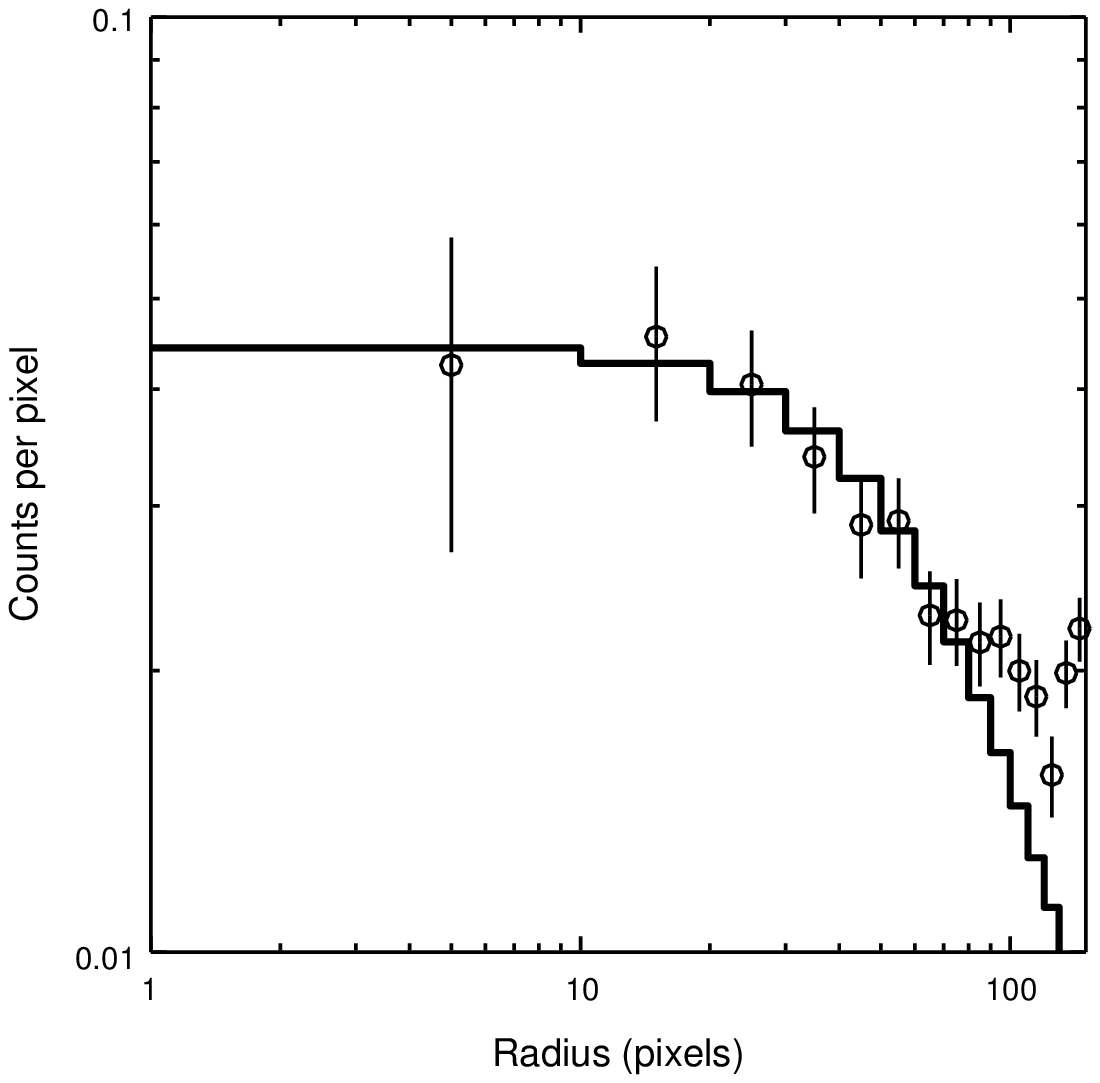}
\includegraphics[width=70mm]{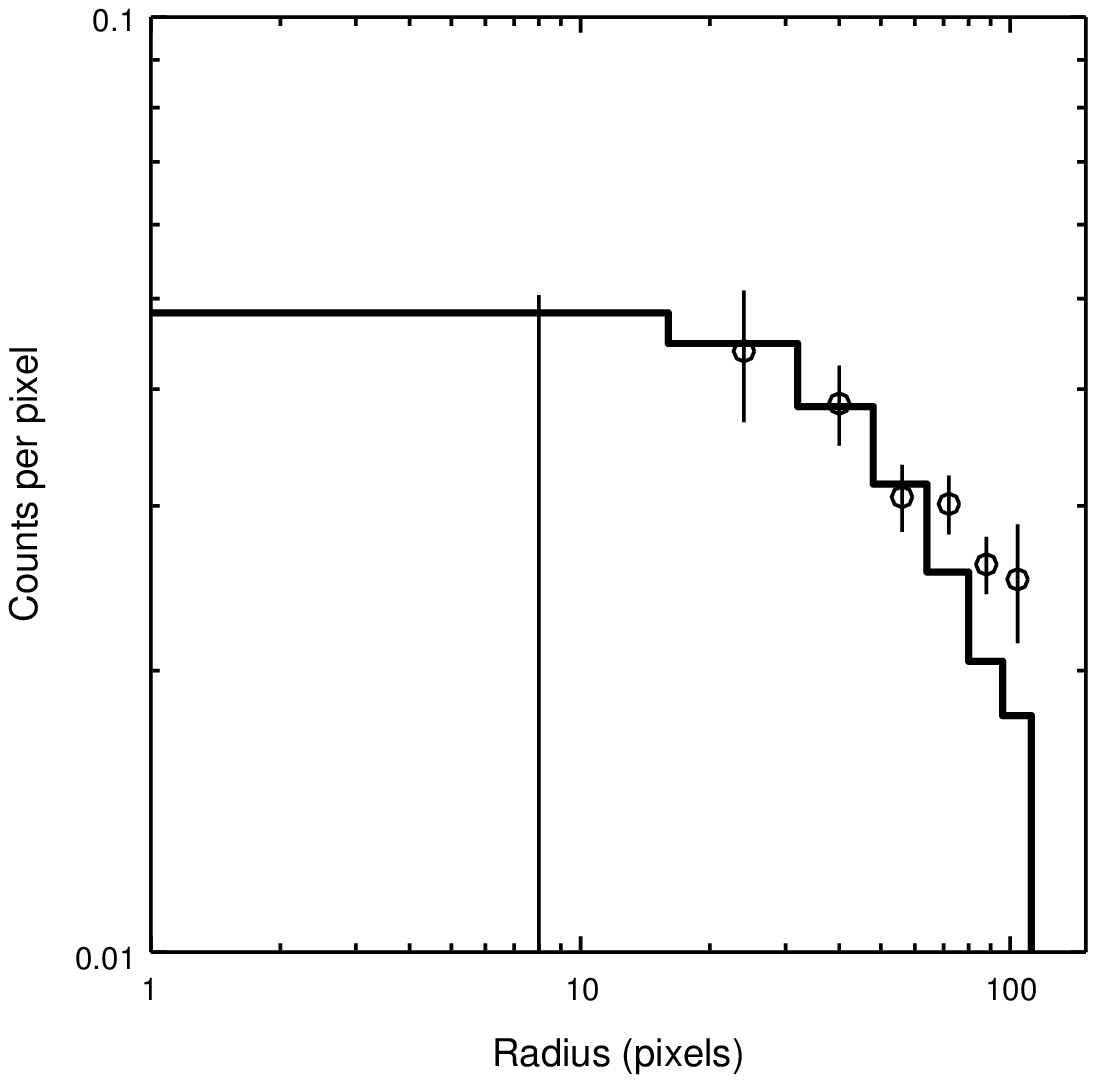}
\includegraphics[width=70mm]{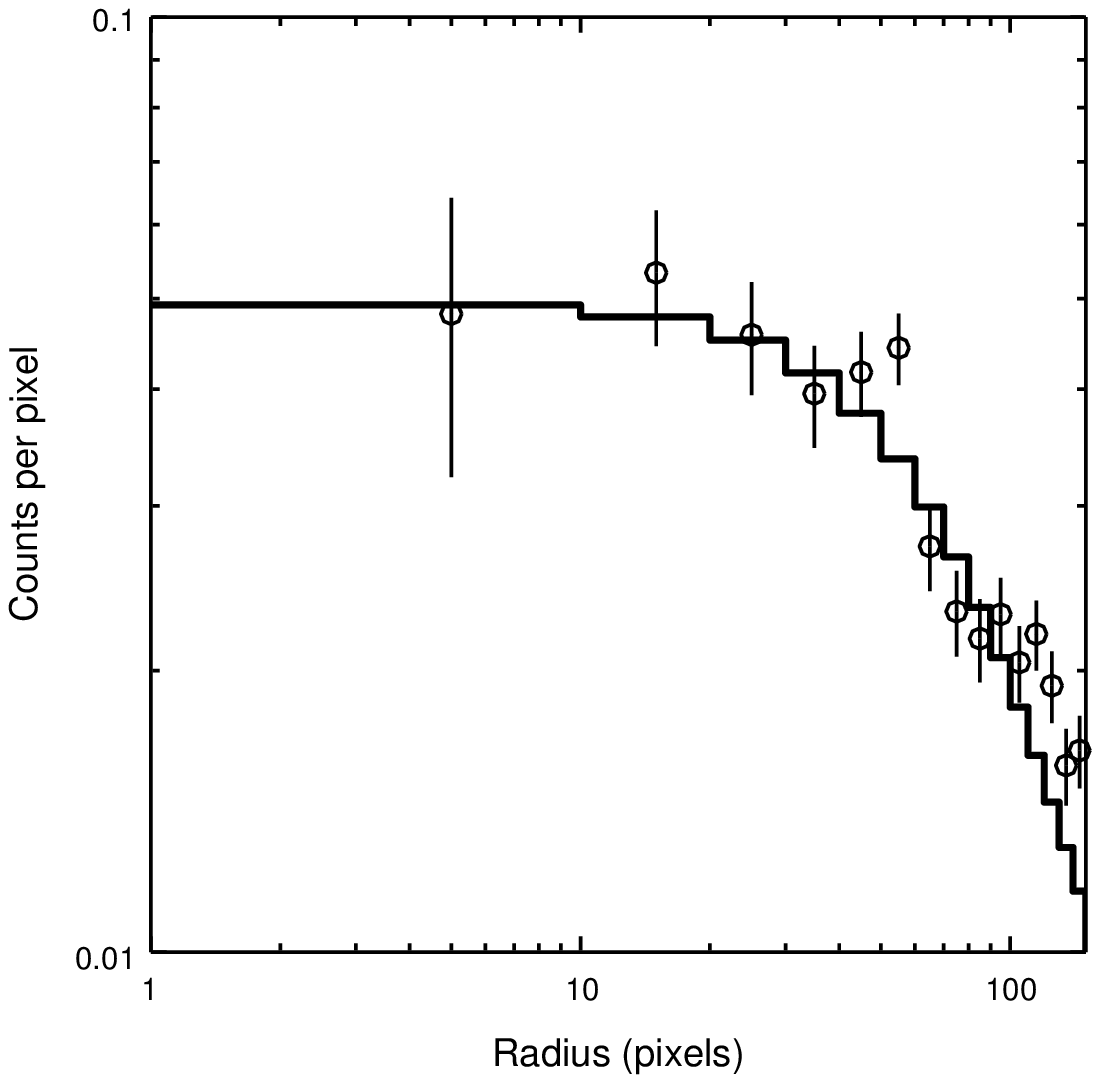}
\caption{ Surface brightness profiles of the low-redshift AEGIS groups, Group 1 (top), Group 2 (middle), and Group 4 (bottom). The data are compared to a beta-model with $\beta = 0.5$ and $r_c = 50$ $h_{70}^{-1}$ kpc convolved with the PSF at the group location.  At a redshift of $z=0.07$, 10 pixels is 6.6 $h_{70}^{-1}$ kpc. }
\end{figure}

For the two low-redshift groups for which we can measure temperatures, Group 1 and Group 4, we find $kT = 0.63$ and $1.5$, respectively.  The third group, Group 2, is the lowest velocity dispersion system in our sample and was previously identified in the SDSS C4 DR3 cluster catalogue (Miller et al.~2005; von der Linden et al.~2007).

\begin{figure}
\includegraphics[width=80mm]{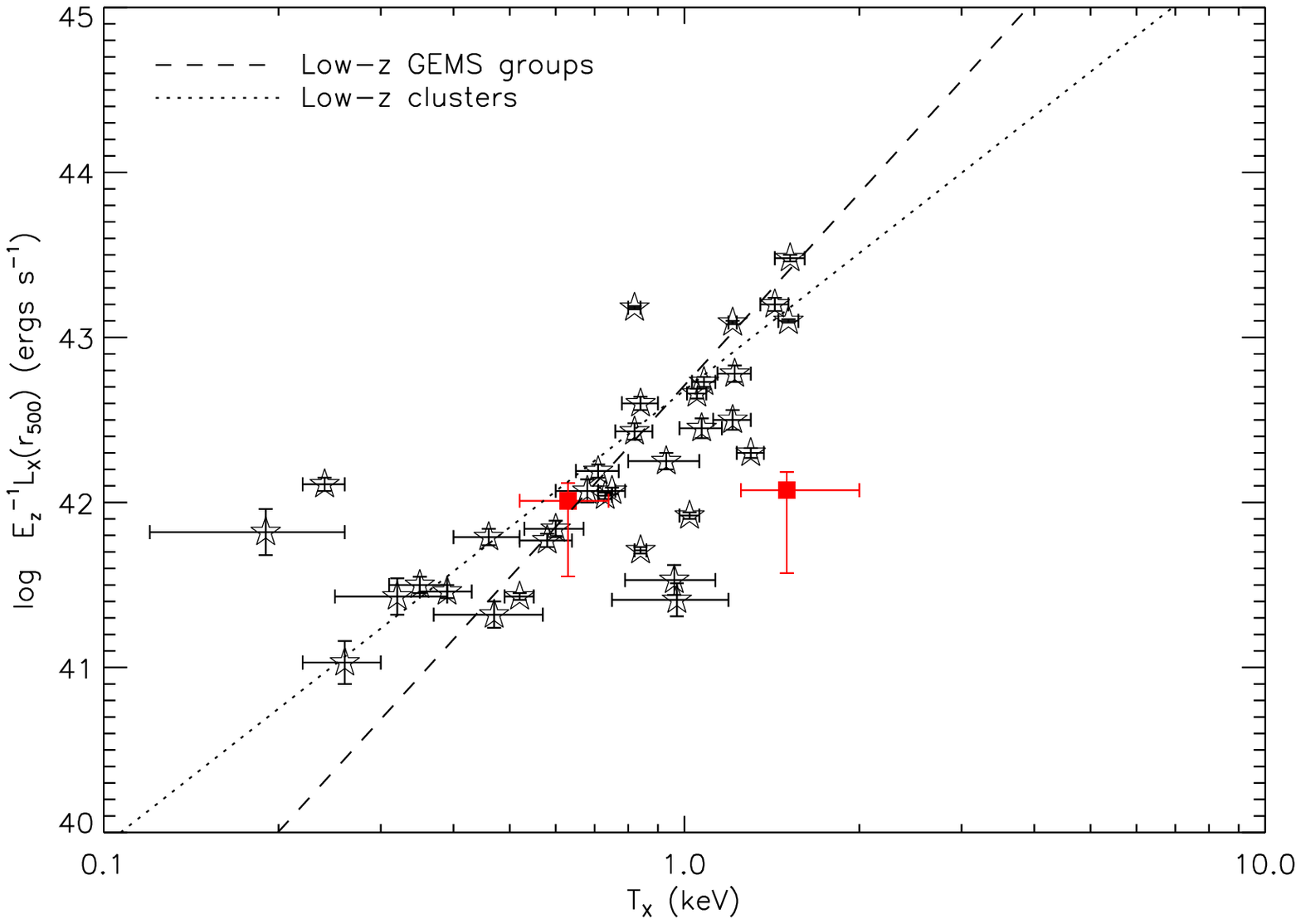}
\includegraphics[width=80mm]{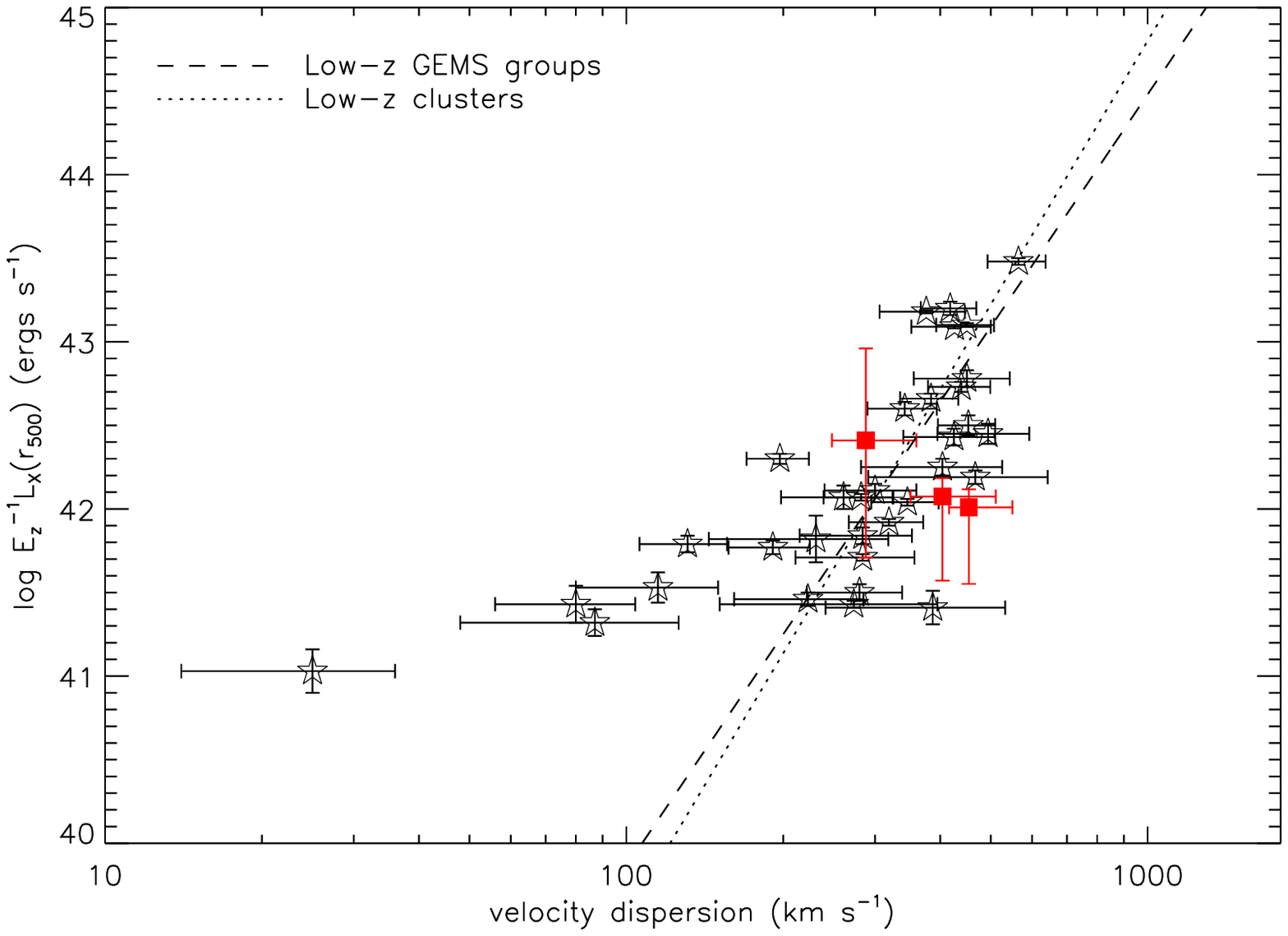}
\includegraphics[width=80mm]{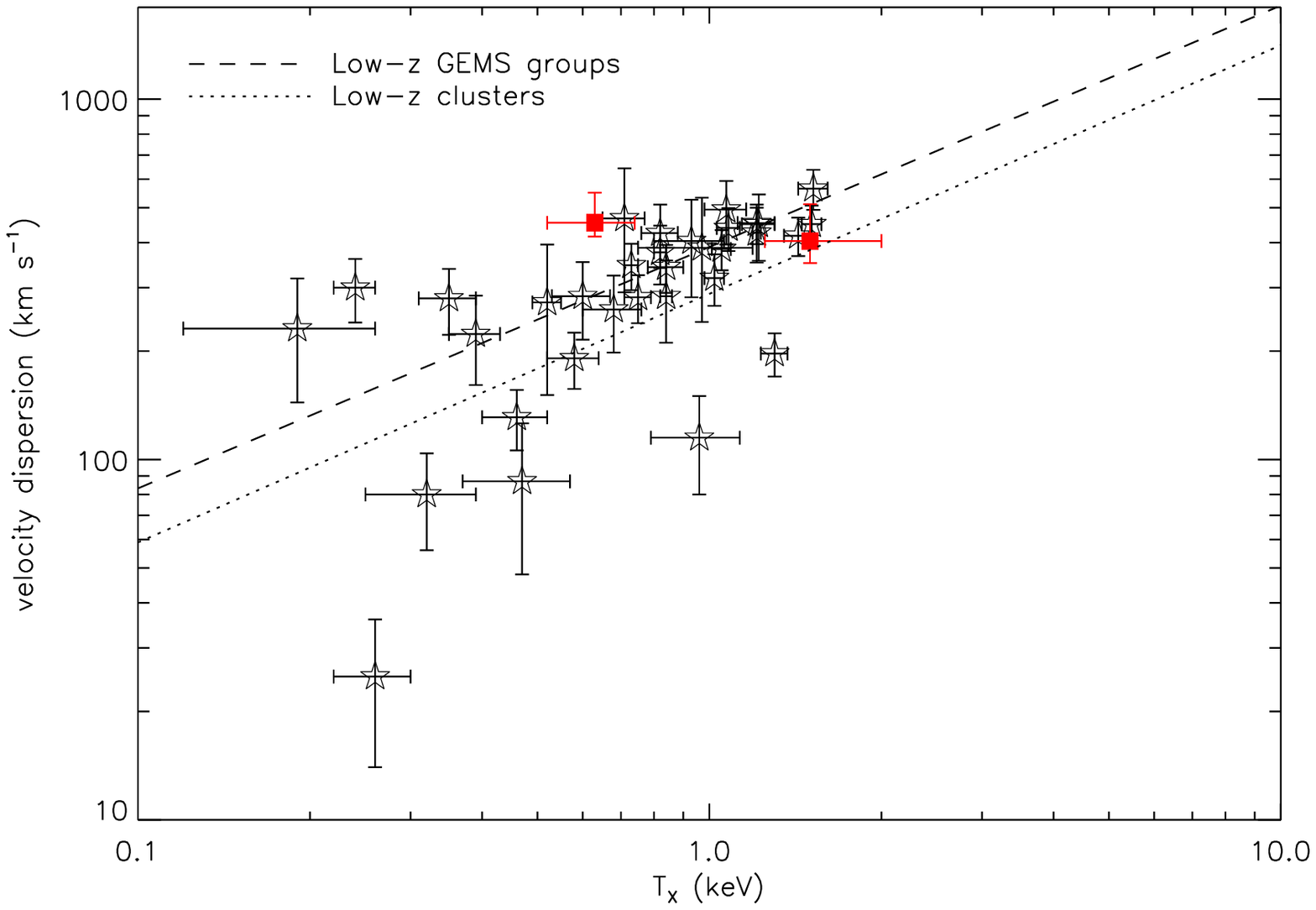}
\caption{ From top to bottom: $L_X-T$, $L_X-\sigma$, and $\sigma-T$ scaling relations for low-redshift groups in AEGIS (red squares) compared to X-ray luminous, low-redshift groups from the GEMS sample (open stars; Osmond \& Ponman 2004).  Fits show the best-fitting relations for the GEMS groups (dashed line; Helsdon \& Ponman, in preparation) and low-redshift clusters (dotted line; Markevitch 1998).  Luminosities for all systems are bolometric (0.01-100 keV) and calculated within $r_{500}$. }
\end{figure}

In Figure 5, we plot the $L_X-T$, $L_X-\sigma$, and $\sigma-T$ scaling relations for the low-redshift AEGIS groups (red squares) compared to the relations observed for other low-redshift groups and clusters with extended X-ray emission \citep{gems,m98}.  Within the uncertainties on the individual group properties and the scatter observed in these scaling relations, the AEGIS groups are consistent with other low-redshift groups.  However, Group 4 appears to have a temperature which is high compared to its X-ray luminosity.  As discussed in \S4.3, this group has a central AGN which could contribute to heating the gas.

\subsubsection{Central Galaxies}

As noted above, the X-ray emission for all three low-redshift groups is centred on a luminous elliptical galaxy, and despite the extended nature of the emission the observed X-ray luminosities are not significantly different from those found for isolated, massive elliptical galaxies.  As an indication of how much of the detected X-ray emission might be due to the central group galaxy, in Figure 6, we compare the detected X-ray luminosities to those of nearby elliptical galaxies.  The X-ray luminosity of galaxies is known to correlate with optical/near-IR luminosity, although with large scatter, indicating a correlation between X-ray emission and galaxy stellar mass.  Here we consider the correlation between $L_X$ and $L_K$, because $K$ band luminosity is more tightly correlated with stellar mass than bluer bands.  $K_s$ band luminosities for the central galaxies are taken from 2MASS.  To compare to galaxy samples from the literature, here we use the X-ray luminosity in the $0.5-2$ keV band, and we do not assume that the X-ray emission extends beyond the detection radius 
(i.e.~we use the luminosity measured within $r_{spec}$ rather than projecting to $r_{500}$).

In Figure 6, we compare the $L_K-L_{X,0.5-2 keV}$ relation for the AEGIS systems (red squares) to a sample of satellite early-type galaxies (central galaxies are excluded) in low-redshift clusters and groups \citep{halo,s07} observed with \textit{Chandra} and to a sample of field early-types observed with \textit{ROSAT} \citep{eo06}.  As noted in \citet{halo}, there appears to be an offset between satellite galaxies in groups and clusters and more isolated ellipticals in this relation with satellite galaxies having lower average X-ray luminosities for their K-band luminosities.  This offset may indicate that some of the hot gas has been stripped from galaxies in groups and clusters, but higher resolution X-ray observations of a field sample are needed to check this result.  We might expect central galaxies to have similar or brighter X-ray emission to field galaxies.  We find, however, that the AEGIS systems span the full range of $L_K-L_{X,0.5-2 keV}$ shown by field, group, and cluster galaxies.  Two of the three low-redshift AEGIS groups have X-ray luminosities which are completely consistent with the K-band luminosities of their central galaxy without the need for group X-ray emission.  One group, Group 1, with the faintest central galaxy may have excess X-ray emission but is also consistent within the observed scatter with the $L_K-L_{X,0.5-2 keV}$ relation of field galaxies.  For this group in particular, we see from Figure 1 that the detected X-ray emission extends beyond the central galaxy and includes other group members.  However, it is clear that the detected X-ray emission for all of the low-redshift AEGIS groups likely has a significant contribution from, or is dominated by, the central galaxy.

\begin{figure}
\includegraphics[width=80mm]{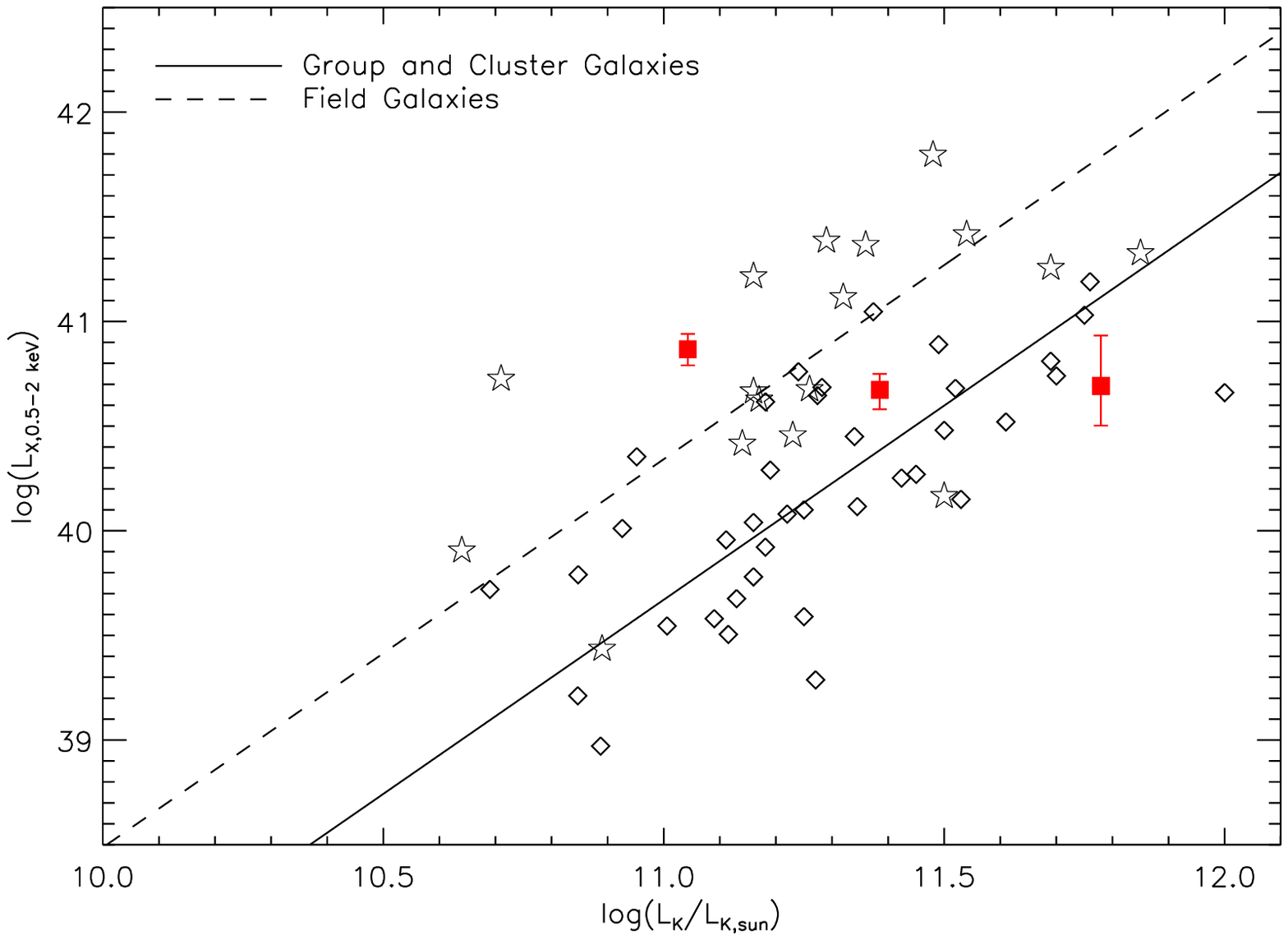}
\caption{ Comparison of the X-ray luminosities of the low-redshift AEGIS groups to local elliptical galaxies.  Specifically, the $L_K-L_{X,0.5-2 keV}$ relation for the AEGIS systems (red squares) is compared to non-central early-type galaxies in groups and clusters (diamonds; Jeltema et al.~2008b, Sun et al.~2007) and to early-type galaxies in the field (stars; Ellis \& O'Sullivan 2006).  The best-fitting relations are shown for the group and cluster galaxies (solid line) and the field galaxies (dashed line).  Here the X-ray luminosities are in the 0.5-2 keV band and are not extrapolated. }
\end{figure}

\subsection{ High-Redshift Groups }

The three high-redshift groups with extended X-ray emission are shown in Figure 2.  These groups are all at redshifts above $z=0.7$.  They represent the first X-ray confirmation of spectroscopically selected DEEP2 groups and have velocity dispersions at the high end of those found for groups in the DEEP2 survey \citep{ge05, ge07}.  A previous search for X-ray emission from DEEP2 groups based the single \textit{Chandra} pointing in the AEGIS region taken in A03 yielded no detections and concluded that the DEEP2 groups appear to be underluminous in X-rays compared to low-redshift groups with similar velocity dispersions \citep{fang}.  However, this one pointing contained relatively few groups (7), all of which had only 3-6 member galaxies making their velocity dispersions uncertain.  In fact, we do not detect any groups in this pointing.  The high-redshift systems detected here have bolometric X-ray luminosities of a few times $10^{43}$ $h_{70}^{-2}$ ergs s$^{-1}$ and temperatures of $2-3.6$ keV, placing them in the massive group to poor cluster regime of galaxy associations.  In particular, the highest redshift system in our sample has a velocity dispersion of $\sim 700$ km s$^{-1}$ and $kT \sim 3.6$ keV; at a redshift of $z=1.13$, it is one of only a few such low-mass systems at $z>1.0$ observed in X-rays \citep{fi07,cdfn}.

\begin{figure}
\includegraphics[width=80mm]{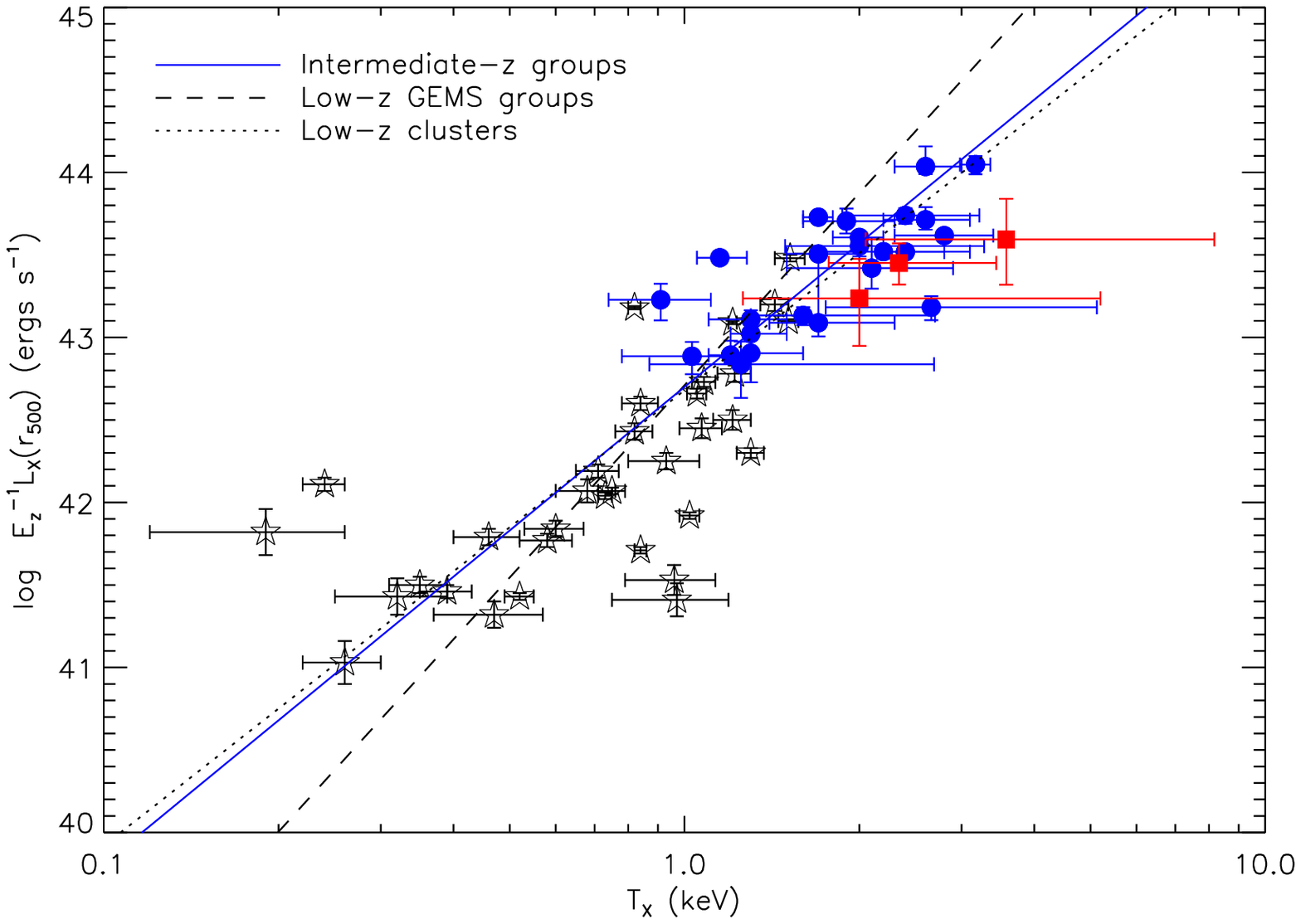}
\includegraphics[width=80mm]{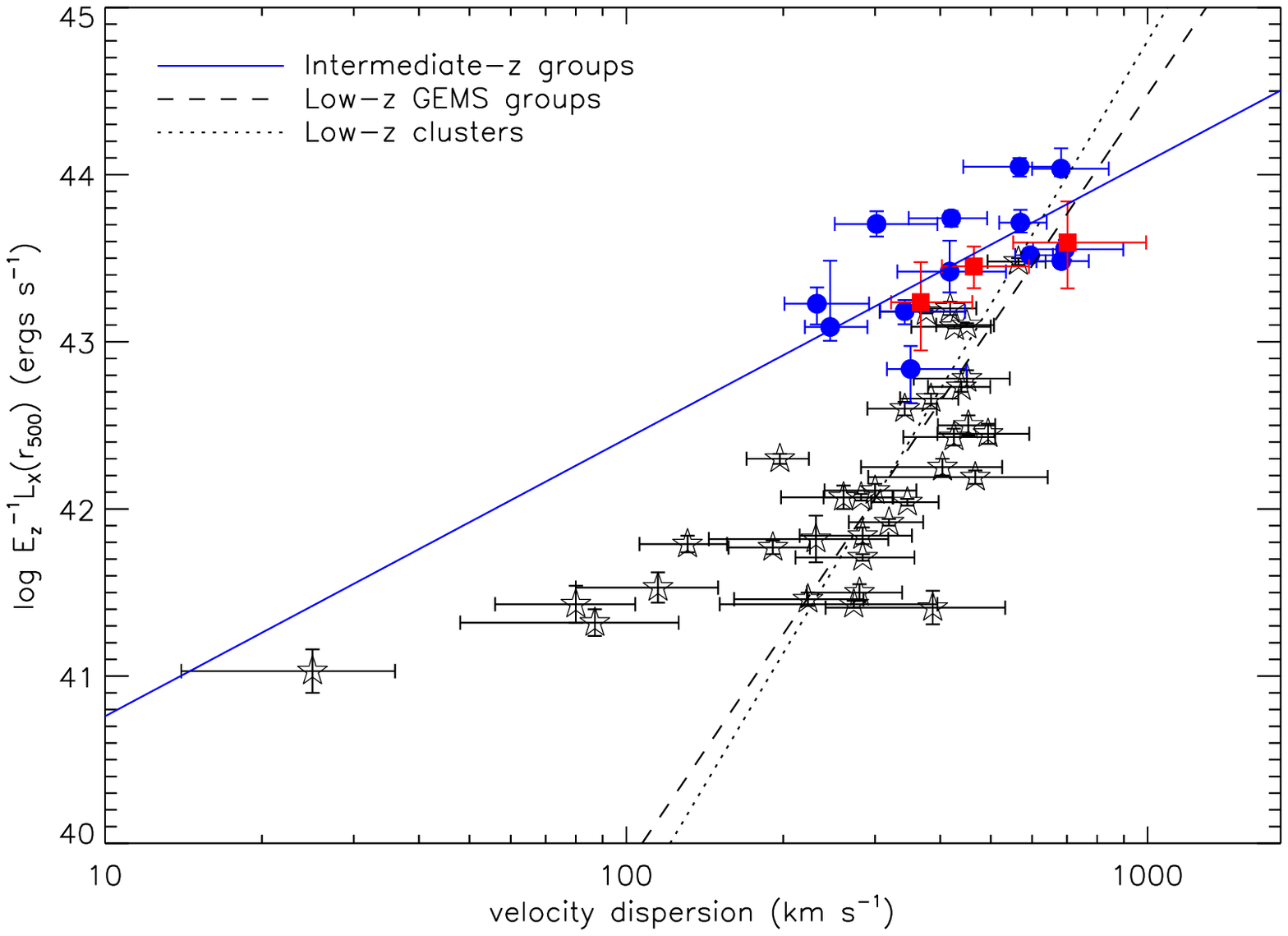}
\includegraphics[width=80mm]{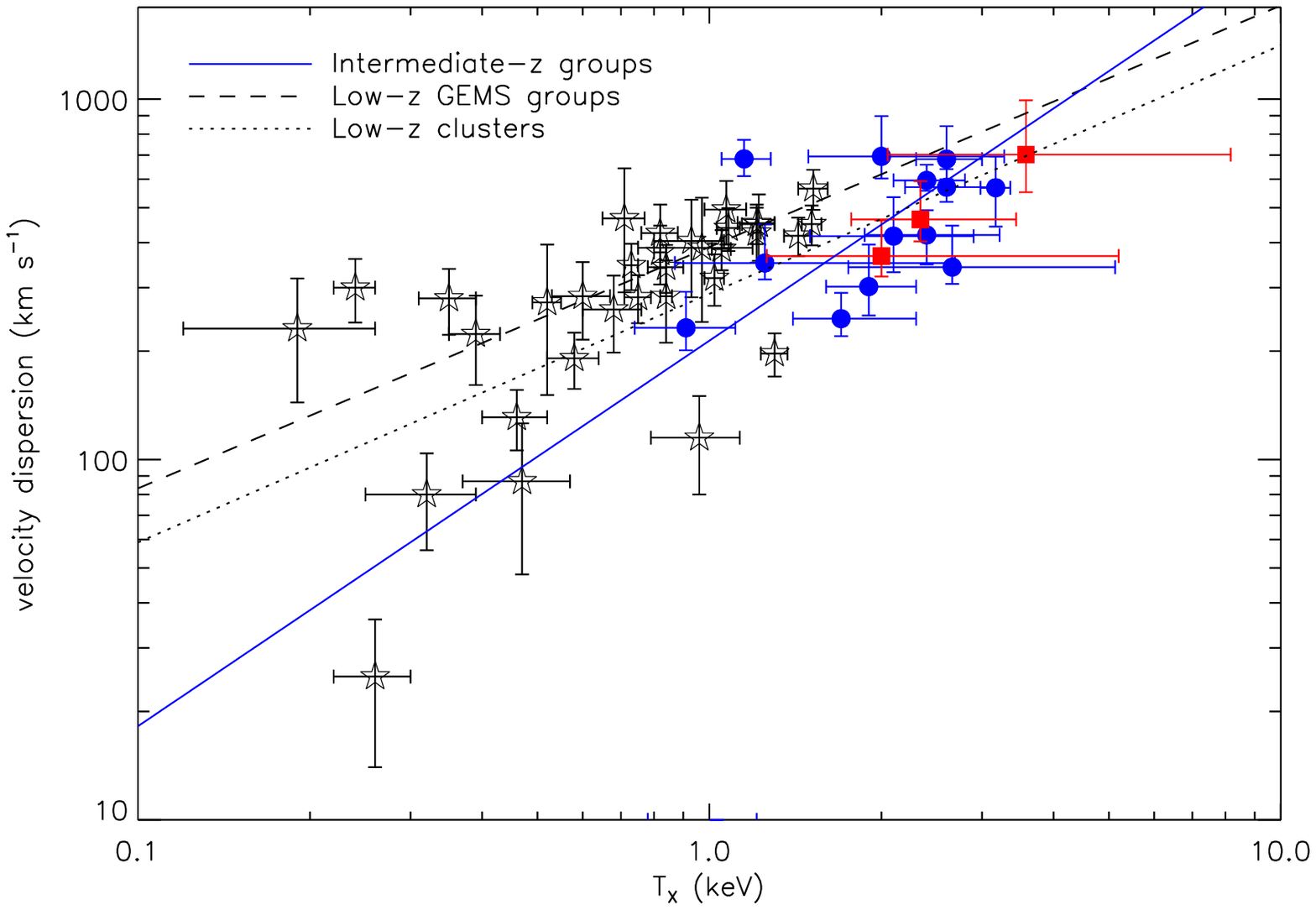}
\caption{ From top to bottom: $L_X-T$, $L_X-\sigma$, and $\sigma-T$ scaling relations for high-redshift groups in AEGIS ($z>0.7$; red squares) compared to X-ray luminous intermediate-redshift groups ($0.2<z<0.6$) from the RDCS and XMM-LSS samples (blue circles; Jeltema et al.~2008a, 2007; Pacaud et al. 2007; Willis et al. 2005) and low-redshift groups from the GEMS sample (open stars; Osmond \& Ponman 2004).  Fits show the best-fitting relations for the intermediate-redshift groups (blue solid line), the low-redshift GEMS groups (dashed line; Helsdon \& Ponman, in preparation) and low-redshift clusters (dotted line; Markevitch 1998).  Luminosities for all systems are bolometric (0.01-100 keV) and calculated within $r_{500}$. }
\end{figure}

Figure 7 shows the $L_X-T$, $L_X-\sigma$, and $\sigma-T$ scaling relations for the high-redshift AEGIS groups.  As in Figure 5, we compare the AEGIS systems to the scaling relations observed for low-redshift groups and clusters.  Here we also show the relations for X-ray luminous, intermediate-redshift groups/poor clusters at $0.2<z<0.6$ derived primarily from two samples: dedicated optical and X-ray follow-up of groups discovered in the \textit{ROSAT} Deep Cluster Survey (RDCS; Jeltema et al.~2008a, 2007,2006; Mulchaey et al.~2006) and the XMM-LSS (Pacaud et al. 2007; Willis et al. 2005).  The best-fitting scaling relations for the intermediate-redshift systems are taken from \citet{1648}.  The intermediate-redshift group sample shows a similar range in X-ray luminosity and temperature to the higher-redshift AEGIS groups.  An $E_z^{-1} = H(z)/H_0$ correction is applied to the $L_X-T$ and $L_X-\sigma$ relations to account for the expected self-similar evolution with redshift of the scaling relations as calculated within a spherical overdensity radius (e.g.~Bryan \& Norman 1998).

The high-redshift AEGIS groups agree quite well with the scaling relations found for lower-redshift systems, with no indications of significant evolution.  They are particularly well-matched to the intermediate-redshift groups, showing a similar flatter relationship in $L_X-\sigma$ and steeper relationship in $\sigma-T$ compared to local groups and clusters.  From these three X-ray selected systems, there is no indication that DEEP2 groups are underluminous in X-rays for their velocity dispersions; however, there are other DEEP2 groups with similar velocity dispersions and redshifts within the \textit{Chandra} coverage that are not found by our X-ray selection.  The X-ray detected group at $z=1.13$ is a more massive system, but taking the two X-ray groups at $z \sim 0.75$ with $\sigma = 350-500$ km s$^{-1}$ as a model, there are five groups not detected in X-rays with $\sigma \geq 350$ km s$^{-1}$, $z \leq 0.8$, and at least five member galaxies in the DEEP2 catalogue within the AEGIS region \citep{ge05}.  Here we limit the sample to groups with at least five members to reduce the uncertainty in the velocity dispersions.  For five members and $\sigma = 350$ km s$^{-1}$, the 1$\sigma$ error range is $249-581$ km s$^{-1}$.  The X-ray properties of optical spectroscopically selected groups, including upper limits and average luminosities through stacking, will be considered in detail in a future paper after the completion of a new AEGIS group catalogue taking advantage of newer spectroscopy and optimised for the AEGIS spectroscopic selection (Gerke et al.~2009, in preparation).  As mentioned in \S3.2, the DEEP2 group selection in the AEGIS region may have a higher false positive rate due to the denser galaxy sampling compared to the rest of the DEEP2 survey.  In addition, deeper X-ray data with a total exposure time of 800 ksecs is currently being taken for $\sim40$\% of the field and will be used in future studies.

\subsubsection{Central Galaxies}

As stated above, at low redshifts X-ray luminous groups are typically found to have bright elliptical galaxies near the X-ray peak.  The same is true of the low-redshift AEGIS groups presented here.  At intermediate-redshifts ($0.2<z<0.6$), follow-up of X-ray-selected groups from the RDCS, found a range of central galaxy properties for systems of similar X-ray luminosity to the high-redshift AEGIS groups \citep{1648,j07, mu06}.  More specifically, only a couple of the RDCS groups have single, dominant central galaxies.  In some groups the central galaxy has multiple luminous nuclei, while other groups have no clear brightest cluster galaxy (BCG), containing instead groups of bright galaxies and/or a brightest galaxy that is significantly offset from the X-ray centre.

\begin{figure}
\begin{center}
\includegraphics[width=60mm]{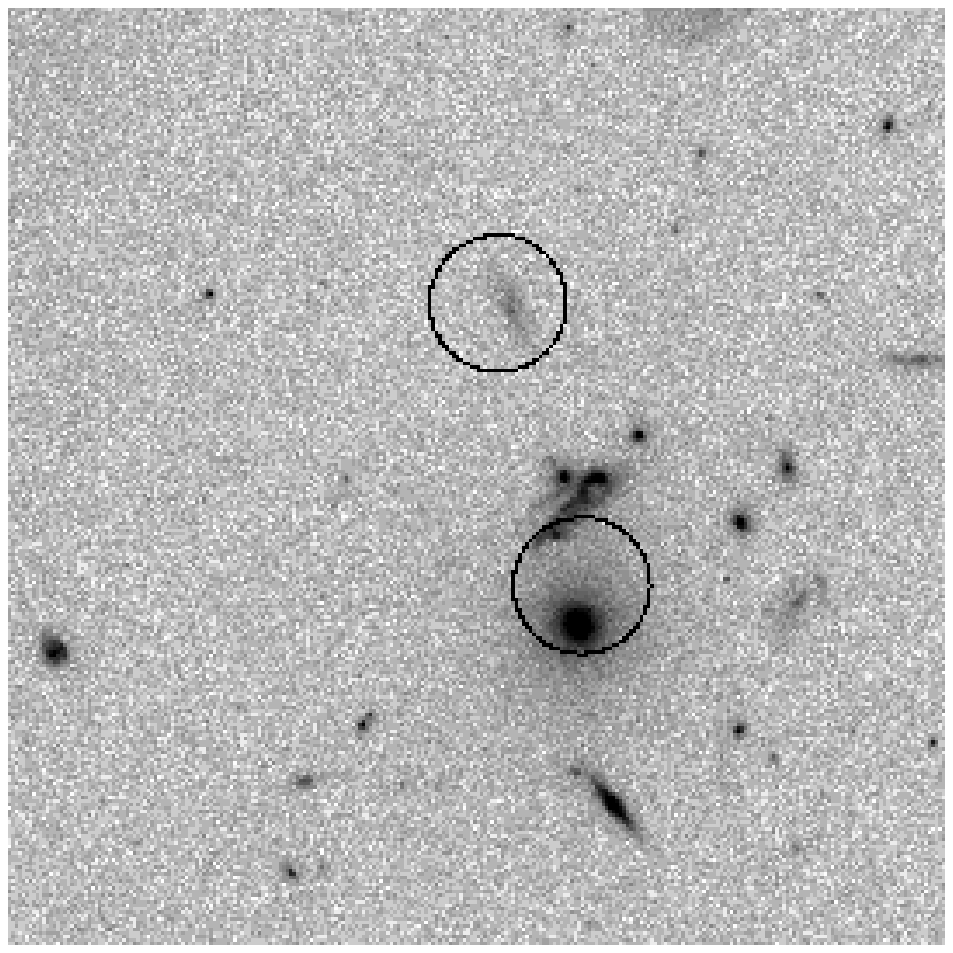}
\includegraphics[width=60mm]{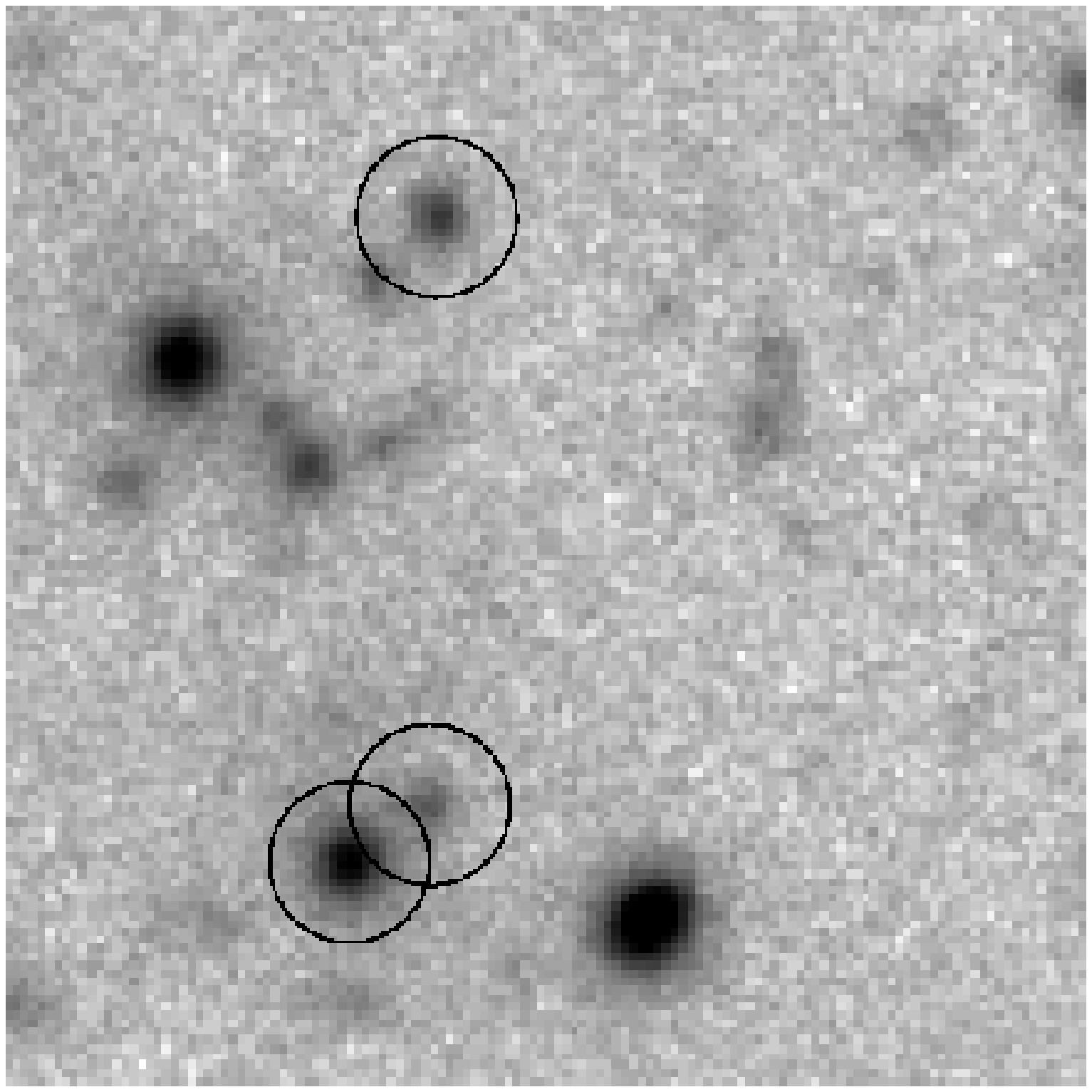}
\includegraphics[width=60mm]{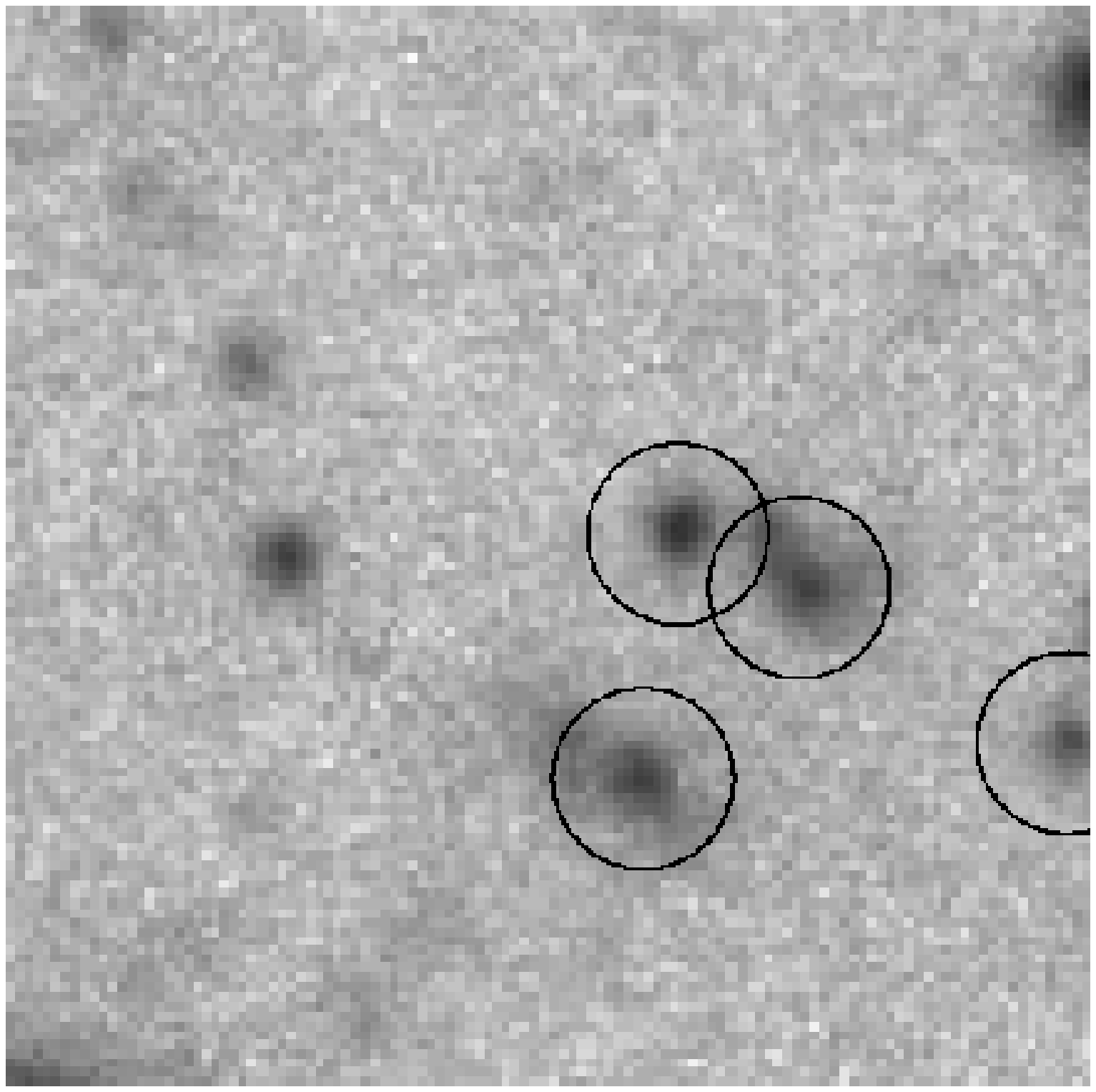}
\end{center}
\caption{ Optical images of the central regions of the high-redshift AEGIS groups, from top to bottom:  Group 5, Group 6, and Group 7.  For Group 5, the HST-ACS i-band image is shown, while for the other groups we show CFHT R-band images.  Images are 200 $h_{70}^{-1}$ kpc on a side and centred on the approximate X-ray centre.  Group member galaxies are marked with black circles.  All of the groups have plausible central galaxies or galaxy pairs. }
\end{figure}

Unfortunately, the peak of the X-ray emission is not well-determined for the faint, high-redshift AEGIS groups, but here we consider whether there are group members which might be central galaxies.  In Figure 8, we show optical images of the group galaxies within 100 $h_{70}^{-1}$ kpc of the X-ray centre as determined by VTPDETECT.  All three high-redshift groups contain potential central galaxies (i.e.~consistent with the X-ray peak).  For Group 5, in particular, ACS imaging is available and shows a bright elliptical group member near the X-ray centre whose colours place it on the red sequence.  The lack of ACS imaging for the other two groups makes the assessment of galaxy morphology more challenging.  Group 6 also appears to have a bright, red sequence, likely early-type galaxy near the X-ray centre; this galaxy has a smaller companion nearby in both position and velocity space.  The highest-redshift system, Group 7, contains a group of similar magnitude galaxies near the centre.  Closest to the centre of this group is a galaxy pair. Both galaxies in the pair fall at the bright end of the green valley in colour-magnitude space, unlike the other central galaxies, and at least one appears asymmetric/disky at the resolution of the CFHT image. We may, therefore, be observing an earlier stage of BCG evolution in the highest redshift group.

Figure 9 shows the position of the likely central group galaxies on the colour-magnitude diagram (left) as well as their spectra (right). While the centrals of Groups 5 and 6 show strong stellar absorption features,
typical of passive galaxies, both galaxies in the central pair of Group 7 show
strong [OII] emission, indicative of ongoing star-formation or AGN activity.  In particular, the spectrum of galaxy 7b also shows weak [NeIII] and [NeV] emission indicating the presence of an AGN in this green valley central galaxy.  

The right panel of Figure 2 shows that for all three groups, the galaxies near the X-ray peak also lie near the centre of the group velocity distribution, as one would expect.  In summary, all of the X-ray detected AEGIS groups have likely central galaxies or galaxy associations.  In two groups, the central galaxy may have one or more companions.  In the highest redshift group at $z=1.13$, the central galaxy pair appears to have younger stellar populations than are typically seen for centrals in X-ray luminous groups at lower redshifts, and one member of this pair hosts an AGN seen through its optical emission lines.

\onecolumn
\begin{figure}
\begin{center}
\includegraphics[width=180mm]{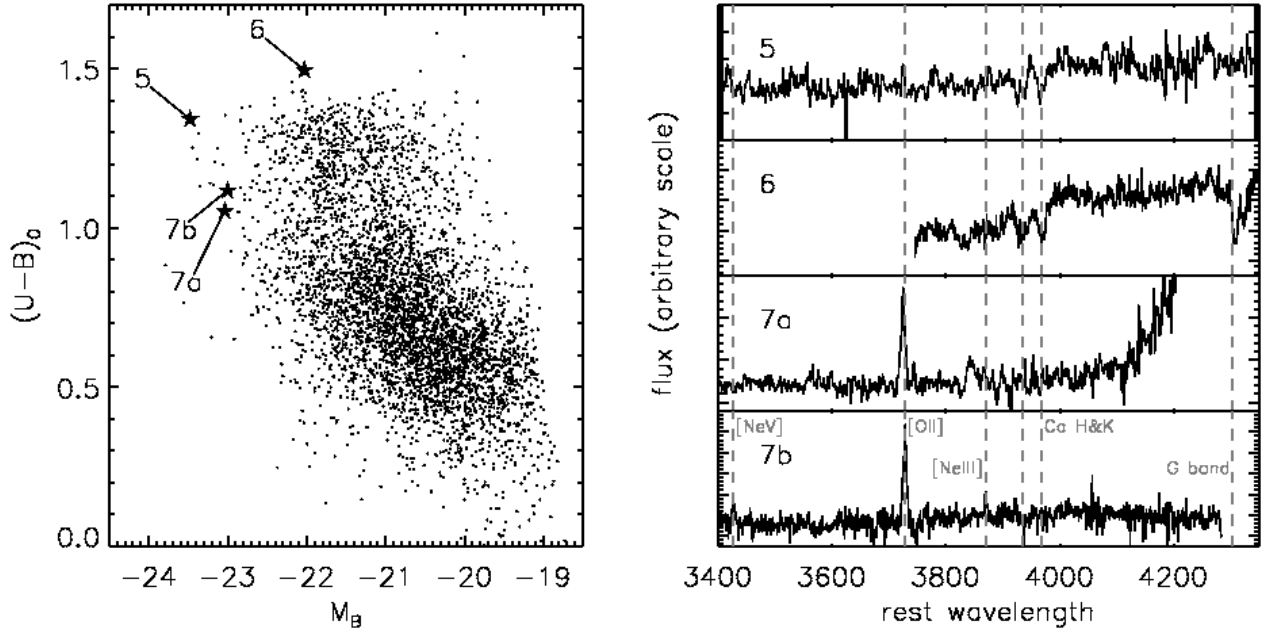}
\end{center}
\caption{Properties of the central galaxies in our high-redshift groups.  We show
properties for the bright central galaxies of Groups 5 and 6 and the two
candidate centrals in Group 7, which we label 7a and 7b.  Left: position of the
central galaxies in colour-magnitude space compared to AEGIS galaxies in the redshift range $0.73<z<1.13$.  Groups 5 and 6 lie on the bright end
of the red sequence, suggesting that these are red-and-dead objects, as typical
of local groups.  The central pair of Group 7, by contrast, lies at the bright
end of the "green valley" between the red and blue populations, suggesting that
these galaxies may not have fully ceased star-formation.  Right: optical spectra
for each of the central galaxies.  Centrals 6 and 7b were observed with the
DEIMOS spectrograph on the Keck II telescope, as part of the DEEP2 Galaxy
Redshift Survey; centrals 5 and 7a were observed with the MMT.  Each spectrum
has been shifted to its rest frame, and prominent spectral features are labelled.
While the centrals of Groups 5 and 6 show strong stellar absorption features,
typical of passive galaxies, both galaxies in the central pair of Group 7 show
strong [OII] emission, indicative of ongoing star-formation or AGN activity.  The presence of both [NeIII] and [NeV] emission in the spectrum of 7b indicate that this galaxy does in fact host an AGN.}
\end{figure}

\twocolumn
\subsection{Group AGN}

As is evident from Figures 1 and 2, in addition to diffuse X-ray emission, we detect X-ray point sources associated with several group member galaxies.  In particular, for all three low-redshift groups we detect point sources associated with the central elliptical galaxy.  These central X-ray sources have relatively low luminosities of $< 10^{42}$ $h_{70}^{-2}$ ergs s$^{-1}$. The central location of these sources within their respective elliptical galaxies and their relatively high X-ray luminosities compared to low mass X-ray binaries (LMXBs), however, make them likely low-luminosity AGN.  The central galaxy in Group 4 shows a pair of X-ray point sources with broad band (0.5-10 keV) luminosities of $2.8 \times 10^{41}$ and $5.4 \times 10^{40}$ $h_{70}^{-2}$ ergs s$^{-1}$, with the latter slightly offset from the centre of the galaxy.  In addition, the 1.4 GHz VLA data shows extended, bright radio emission associated to the central galaxy, so this group almost certainly hosts a central AGN or dual AGN, given the pair of X-ray point sources.  Figure 10 shows the central galaxy overlaid with contours of radio (red), diffuse X-ray (black), and X-ray point source (green) emission.  The radio emission is extended but not aligned with the optical axis, and a large, faint, asymmetric radio structure appears to the southeast along the jet axis.  From the AEGIS20 radio catalogue, the central radio emission has a luminosity of $1.4 \times 10^{23}$ W Hz$^{-1}$ \citep{iv07}. The MMT spectrum of this galaxy does not show significant [OIII] emission, but does show weak H$\alpha$ emission offset by $\sim 870$ km s$^{-1}$ from the galaxy.  Interestingly, this group has a relatively high X-ray temperature compared to the $L_X-T$ relation.

\begin{figure}
\begin{center}
\includegraphics[width=80mm]{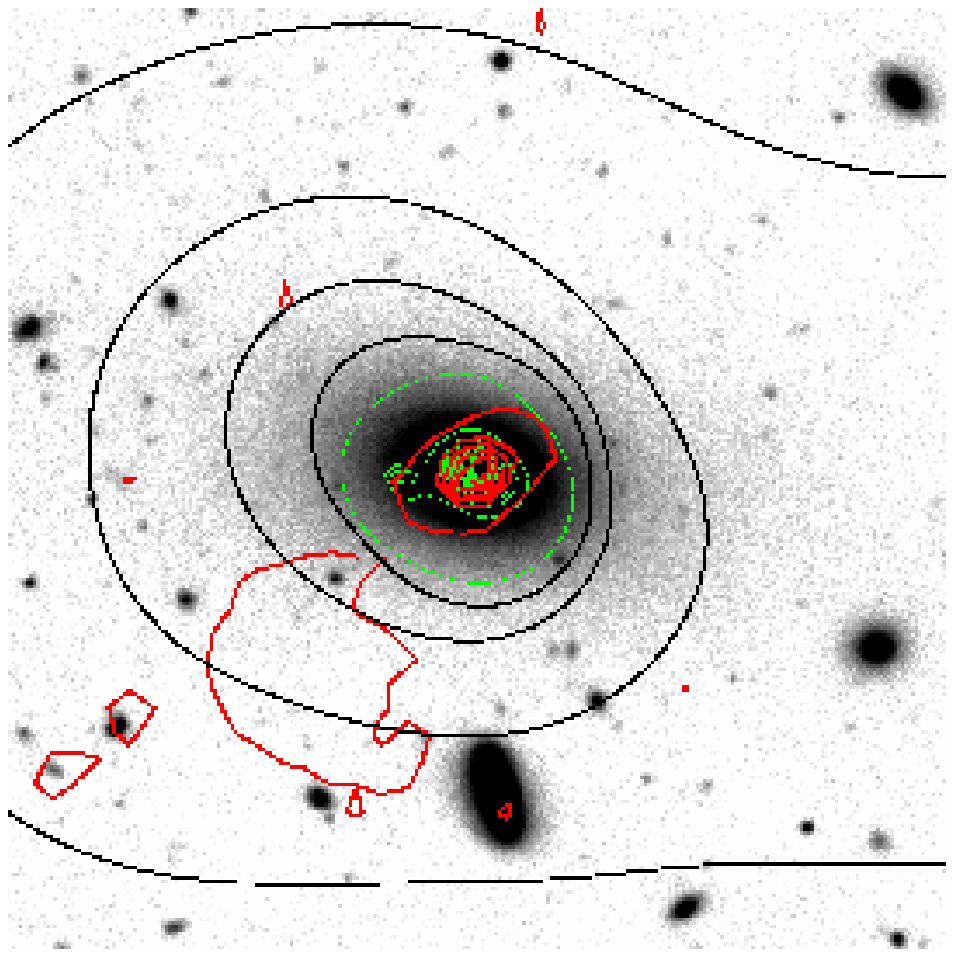}
\includegraphics[width=80mm]{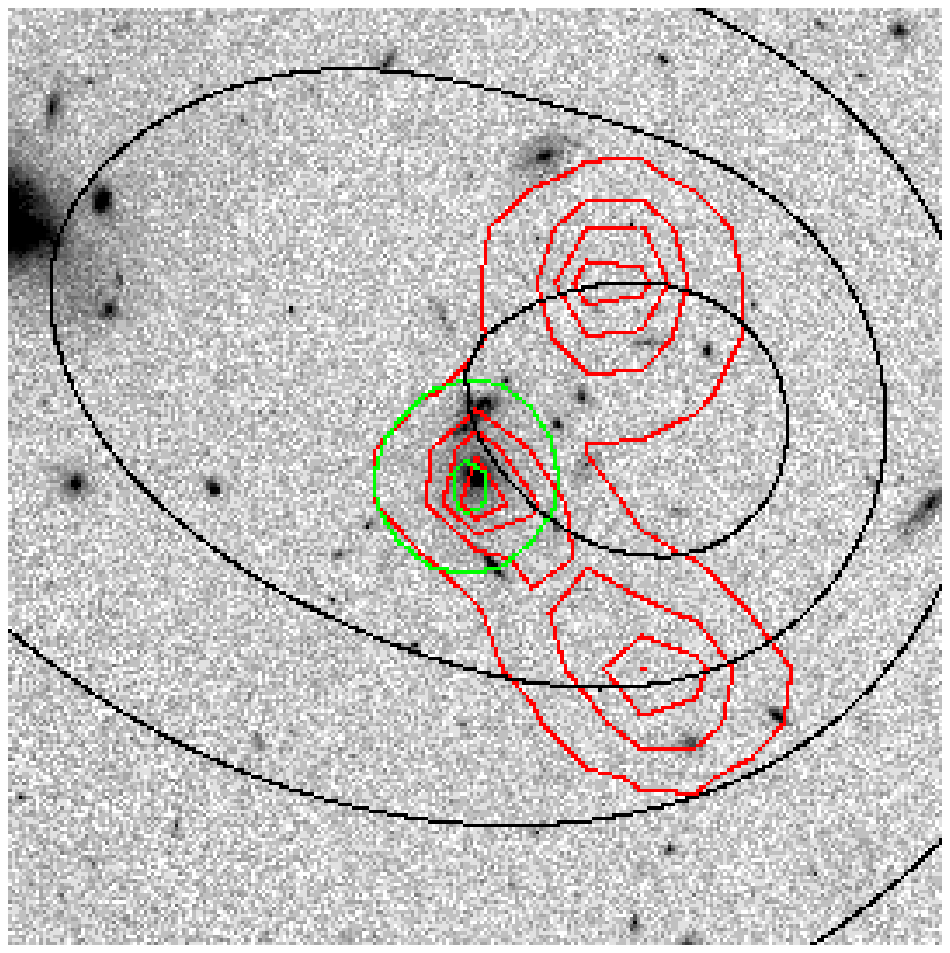}
\end{center}
\caption{Top: CFHT R-band image of the central galaxy in Group 4 overlaid in red with contours of 1.4 GHz radio emission from the VLA.  The image is 150 $h_{70}^{-1}$ kpc on a side.  As in Figure 1, contours of diffuse X-ray emission are shown in black, and X-ray point sources are shown in green. All contours are linearly spaced.  Bottom: Same for the high-redshift group Group 5.  Here the HST-ACS i-band image is shown and the scale is 400 $h_{70}^{-1}$ kpc on a side.}
\end{figure}

The central source in Group 2 has a luminosity of $2.1 \times 10^{41}$ $h_{70}^{-2}$ ergs s$^{-1}$, while the central source in Group 1 has a somewhat lower luminosity of $2.3 \times 10^{40}$ $h_{70}^{-2}$ ergs s$^{-1}$.  Unfortunately, radio data are not available for Groups 1 and 2.  Spitzer data for the BCG in Group 1 show a potential, though not conclusive, excess at 8 $\mu$m and 24 $\mu$m. In addition to the central sources in the low-redshift groups, we detect three X-ray point sources with lower luminosities of $1-3 \times 10^{40}$ $h_{70}^{-2}$ ergs s$^{-1}$ associated with spiral galaxies showing spectral signatures of ongoing star formation.

In the high-redshift groups, we detect two X-ray point sources associated with group member galaxies in the $z \sim 0.75$ groups and one additional X-ray source within the redshift range of Group 5 but outside the group radius.  The two group sources have broad band X-ray luminosities greater than $10^{43}$ $h_{70}^{-2}$ ergs s$^{-1}$ and the neighbouring source has $L_X = 4.5 \times 10^{42}$ $h_{70}^{-2}$ ergs s$^{-1}$, making them difficult to explain as anything other than AGN.  Overall, we find an X-ray AGN fraction in $z \sim 0.75$ groups of $7.4^{+9.8}_{-4.8}$\% for galaxies brighter than $M_R = -20$ with $L_X > 10^{43}$ $h_{70}^{-2}$ ergs s$^{-1}$ (2 out of 27), consistent within the sizeable errors with the cluster fractions of $2.0$\% ($M_R < -20$ and $L_X > 10^{42}$ ergs s$^{-1}$) found at $z \sim 0.6$ by Eastman et al.~(2007) and $2.7$\% ($M_R < -20$ and $L_X > 10^{43}$ ergs s$^{-1}$) found by Kocevski et al.~(2008) at $z \sim 0.9$.  Our AGN fraction is marginally higher than the cluster fraction of $1.2$\% at low redshift (Sivakoff et al.~2008).  A previous study in the AEGIS field of X-ray AGN associated with high-redshift, optically selected DEEP2 groups found an excess of AGN in groups, but once the host galaxy properties were accounted for, the X-ray AGN fraction in groups was consistent with the field (Georgakakis et al.~2008).

One of the high-redshift X-ray point sources is associated with the likely central elliptical galaxy in Group 5.  This galaxy also hosts a bright, bent double radio source with $L_{1.4 GHz} = 5.1 \times 10^{25}$ W Hz$^{-1}$ shown in Figure 10.  The bent nature of this source may indicate that it is moving with respect to the ICM (e.g.~Owen \& Rudnick 1976); however its line-of-sight velocity relative to the group is small.  Alternatively, it has been suggested that bent radio sources may indicate a recent group/cluster merger (e.g.~Burns et al.~2002).  In fact, the galaxy distribution surrounding Group 5 shows that it may be part of a larger-scale/filamentary structure, and there is a potential second peak in the velocity distribution at slightly higher redshift, supporting the possibility of a merger.  Fainter radio sources ($L_{1.4 GHz}$ between $1.5 \times 10^{23}$ W Hz$^{-1}$ and $1.5 \times 10^{24}$  W Hz$^{-1}$) are also found in the other two high-redshift groups associated to the central galaxy in Group 6, the X-ray AGN in Group 6, and both the central pair and the galaxy south of this pair in Group 7.  As noted above, one the the central galaxies in Group 7, 7b, also shows optical emission lines indicating an AGN.

In summary, all of the central group galaxies, and several other groups members, show signs of activity in X-ray, radio, and/or optical emission.  In particular, the central galaxies in Groups 4 and 5 host both X-ray and radio jet sources.

\section{ SUMMARY }

We have conducted an X-ray search for groups of galaxies in the Extended Groth Strip utilising the Chandra coverage of $\sim0.67$ deg$^2$, part of the AEGIS survey.  Groups of galaxies are selected based on the presence of extended X-ray emission and then matched against optically selected groups from the DEEP2 spectroscopy in the field.  In total, we find seven extended X-ray sources, and we identify optical counterparts for six.  Three of these groups lie at $z>0.7$ and represent the first X-ray detections of high-redshift DEEP2 groups.  The other three groups are low-luminosity, low-redshift groups which form part of a larger structure or "supergroup" at $z \sim 0.07$ in the southern portion of the AEGIS field.  The seventh group candidate unfortunately lies outside of the current spectroscopy, but as discussed in the appendix, it is also a high-redshift group candidate.  Our main results are summarised below.

\textit{Low-redshift groups:}  The three low-redshift extended X-ray sources are all centred on luminous elliptical galaxies in groups at $z \sim 0.07$, one of which was previously identified in the SDSS C4 Cluster catalogue.  Together the DEEP2, MMT, and SDSS spectroscopy reveal a large-scale structure in the southern portion of the AEGIS field dominated by these three groups.  The groups are relatively low-mass systems with $kT = 0.6-1.5$ keV and $\sigma = 280-450$ km s$^{-1}$.  When projected to $r_{500}$, their X-ray luminosities are consistent with the total group velocity dispersion; however, their detected X-ray luminosities are similar to those found for individual massive elliptical galaxies.

\textit{High-redshift groups:}  For the high-redshift systems, we find X-ray temperatures in the range of $2-3.6$ keV and bolometric X-ray luminosities of a few times $10^{43}$ $h_{70}^{-2}$ ergs s$^{-1}$, placing them in the massive group to poor cluster regime of galaxy associations.  The hottest system with $kT=3.6^{+4.6}_{-1.5}$ keV lies at a redshift of 1.13, making it one of only a few systems of its mass known at $z>1$.  The extensive DEEP2 spectroscopy supplemented by additional redshifts from MMT allow us to identify $8-20$ members for each group and to determine group velocity dispersions.  These data allow us the first look at the $L_X-T$, $L_X-\sigma$, and $\sigma-T$ scaling relations for systems in this mass range at high redshifts.  We find no evidence for evolution in these scaling relations when comparing the X-ray-selected DEEP2 groups to X-ray luminous groups at lower redshifts.  However, there are X-ray undetected DEEP2 groups in the field with similar velocity dispersions, which will be studied in detail in a future paper.  Similar to X-ray luminous, low-redshift groups, all of the detected systems have potential central galaxies, though two of the central galaxies appear to have close companions.  The likely central galaxies in the two groups at $z\sim0.75$ are bright, red galaxies with strong stellar absorption features typical of passive BCGs.  The central galaxy pair in the highest redshift system at $z=1.13$, however, both show [OII] emission and lie at the bright end of the green valley in colour-magnitude space.  Here we may be witnessing an early stage of BCG formation.

\textit{Group AGN:}  We find X-ray point sources associated to all of the central galaxies in the low-redshift groups and one of the centrals in the high-redshift groups.  We also detect radio sources near the centres of all of the groups with radio observations (all three high-redshift and one low-redshift group).  Particularly convincing and intriguing examples of central AGN activity include the bright bent-double radio source and X-ray point source at the center of Group 5 at $z=0.74$, the extended radio and double X-ray point sources associated to the central galaxy in the lowest-redshift group, Group 4, and the bright green valley pair of galaxies in Group 7 at $z=1.13$, one of which shows optical AGN emission lines.  Considering all group galaxies, we find an X-ray-selected AGN fraction in $z \sim 0.75$ groups of $7.4^{+9.8}_{-4.8}$\% for galaxies brighter than $M_R = -20$ with $L_X > 10^{43}$ $h_{70}^{-2}$ ergs s$^{-1}$, consistent with previous studies of high-redshift clusters \citep{ea07,ko08}.
\\

We would like to sincerely thank the anonymous referee for their comments which led to several improvements in the paper.  We would also like to thank members of the AEGIS team, particularly D.~Rosario and Rob Ivison, for their advice and support with the multiwavelength data.  Support for this work was provided by the National Aeronautics and Space Administration through \textit{Chandra} Award Number AR9-0017X issued by the \textit{Chandra} X-ray Observatory centre, which is operated by the Smithsonian Astrophysical Observatory for and on behalf of the National Aeronautics Space Administration under contract NAS8-03060.  T.E.J. is grateful for support from the Alexander F. Morrison Fellowship, administered through the University of California Observatories and the Regents of the University of California.  B.F.G. was supported by the U.S. Department of Energy under contract number DE-AC3-76SF00515.  Support for M.C.C. was provided by NASA through the {\it Spitzer Space Telescope} Fellowship Program.  E.S.L. acknowledges financial support from the UK Science and Technology Facilities Council.  This study makes use of data from AEGIS, a multiwavelength sky survey conducted with the Chandra, GALEX, Hubble, Keck, CFHT, MMT, Subaru, Palomar, Spitzer, VLA, and other telescopes and supported in part by the NSF, NASA, and the STFC.

\appendix
%\section{ NOTES ON INDIVIDUAL SYSTEMS }
\section{CXOU J141616.6+520601}

As discussed in Section 2.2, one of our extended X-ray group candidates, Group 3, could not be identified optically, because it falls outside the DEEP2 spectroscopic region.  A NED search also reveals no known galaxy redshifts within a couple of arcminutes of the group, though interestingly there is a radio source about $0.3'$ from the X-ray centre.  Here we consider this extended X-ray source in more detail.  The top panel of Figure A1 shows the \textit{Chandra} 0.5-2 keV contours overlaid on the CFHT R-band image.  There are two galaxies lying very close to the peak of the X-ray emission.  Using the CFHT BRI photometry, we estimate photometric redshifts for these galaxies and find $z = 0.05$ ($z<0.2$) for the brighter galaxy and $z=0.9 \pm 0.1$ for the fainter galaxy.

As mentioned in \S3.3, we are able to estimate a temperature for this group of $2.9^{+1.0}_{-0.5}$ keV by leaving the redshift as a free parameter.  This temperature implies a massive group/poor cluster size system.  In fact, this system is one of the hottest and highest flux sources in our sample.  In principle, requiring that the group lie on the $L_X-T$ relation can constrain its redshift, and we illustrate this comparison in the bottom panel of Figure A1.  Here we refit the spectrum and determine the temperature and luminosity for four assumed, fixed redshifts, $z=0.05, 0.25, 0.5,$ and $1.0$.  The luminosities are projected to $r_{500}$ for the assumed redshift as described in \S3.3.  From Figure A1, the $L_X-T$ relation for this group favours a moderate to high redshift, with $z=0.5$ and $z=1.0$ best matching the relation.  However, within the temperature errors and the scatter in the $L_X-T$ relation a wide range of redshifts are possible.

\begin{figure}
\begin{center}
\includegraphics[width=60mm]{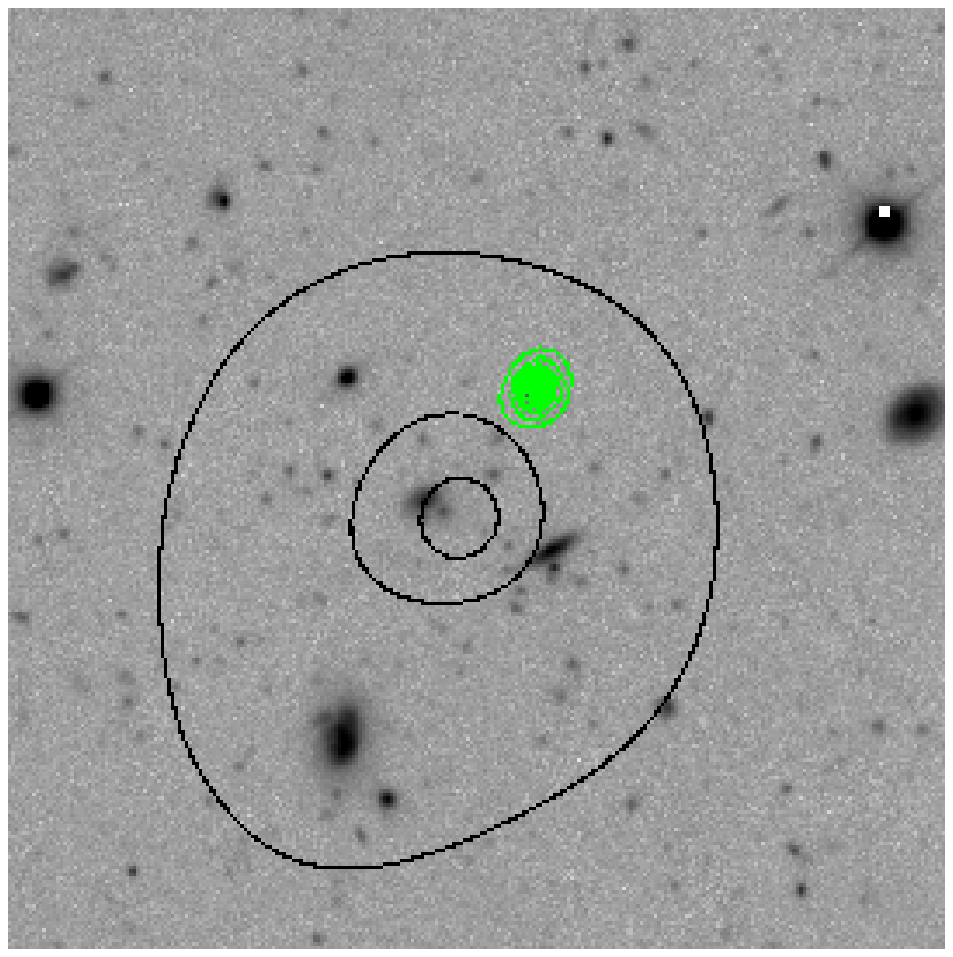}
\includegraphics[width=80mm]{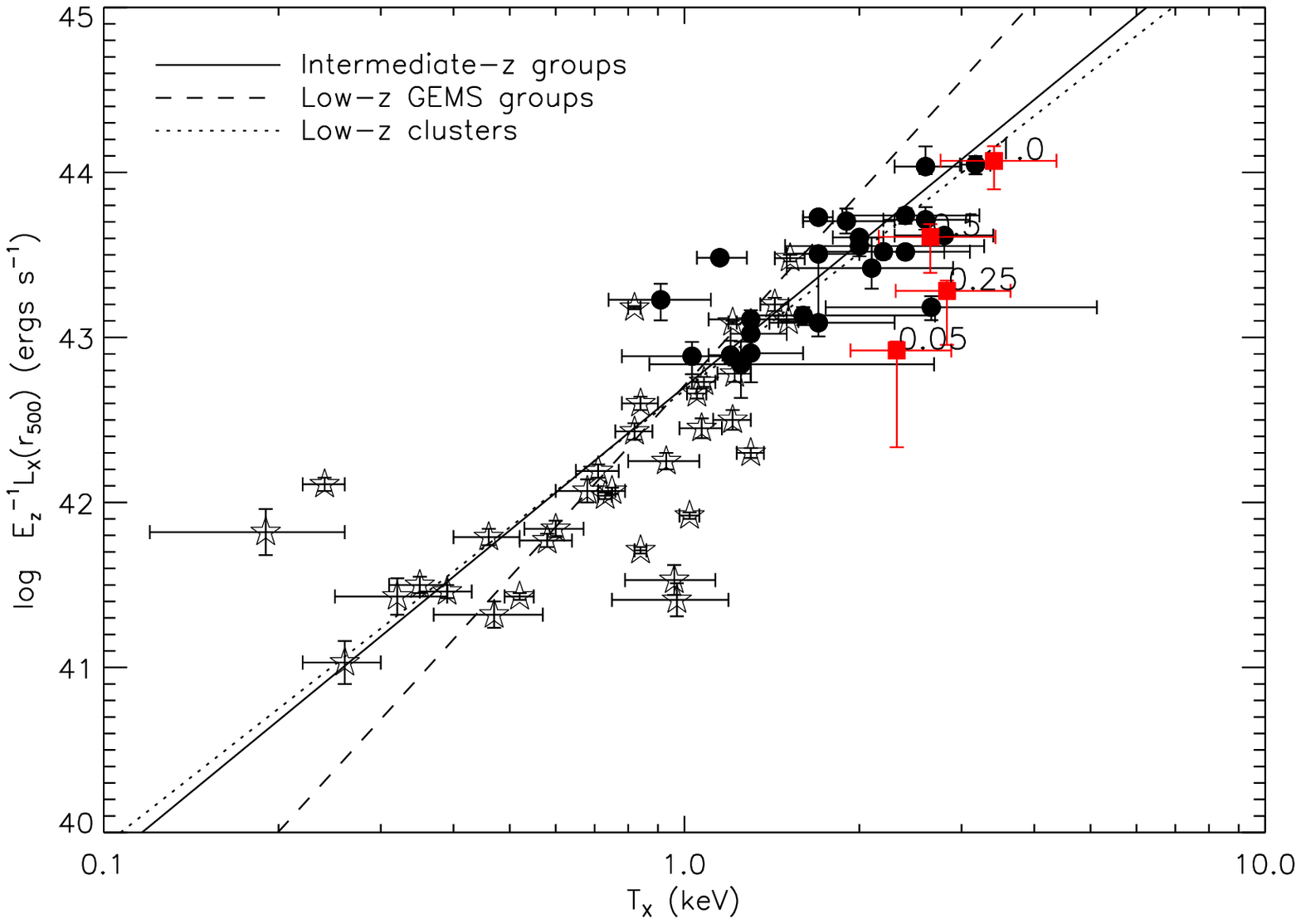}
\end{center}
\caption{ Extended X-ray source CXOU J141616.6+520601 (Group 3).  This source lies outside of the DEEP2 spectroscopic coverage, so no optical identification was possible.  Top:  Contours of diffuse X-ray emission (0.5-2 keV), from an adaptively smoothed image with point sources removed, are shown in black overlaid on the CFHT R-band image.  X-ray point sources are shown in green.  Bottom: $L_X-T$ relation for  Group 3 for different assumed redshifts, $z=0.05, 0.25, 0.5, 1.0$  (red squares) compared to the group and cluster samples shown in Figure 7.  Luminosities for all systems are bolometric (0.01-100 keV) and calculated within $r_{500}$. }
\end{figure}

\end{document}